\catcode`\@=11
\def\gsim{\ifmmode{\mathrel{\mathpalette\@versim>}}
    \else{$\mathrel{\mathpalette\@versim>}$}\fi}
\def\lsim{\ifmmode{\mathrel{\mathpalette\@versim<}}
    \else{$\mathrel{\mathpalette\@versim<}$}\fi}
\def\@versim#1#2{\lower 2.9truept \vbox{\baselineskip 0pt \lineskip 
    0.5truept \ialign{$\m@th#1\hfil##\hfil$\crcr#2\crcr\sim\crcr}}}
\catcode`\@=12
\def\als{\alpha_*}
\def\asn{\alpha_{\rm SN}}
\def\brem{bremsstrahlung$\;\,$}

\def\cx{C_{\rm X}}

\def\ch{C_{\rm HI}}
\def\che{C_{\rm HeI}}
\def\chee{C_{\rm HeII}}
\def\eb{E_{\rm Br}}
\def\eco{E_{\rm C}}
\def\ega{E_{\gamma}}
\def\eph{E_{\rm ph}}

\def\eps{\epsilon}

\def\fz{f_\circ}
\def\fznu{\fz (\nu)}
\def\fb{f_{\rm Br}}
\def\fle{{\cal F}_{\rm BH}}
\def\fBH{f_{\rm BH}}
\def\gef{g_{\rm eff}}
\def\kb{k_{\rm B}}
\def\lb{L_{\rm B}}
\def\lbh{L_{\rm BH}}
\def\lbr{L_{\rm Br}}

\def\ledd{L_{\rm Edd}}

\def\lgm{L_{\rm grav}}
\def\lsn{L_{\rm SN}}
\def\lrad{L_{\rm r}}
\def\lsol{L_\odot}
\def\lx{L_{\rm X}}
\def\lh{L_{\rm h}}
\def\mast{M_*}
\def\mgas{M_{\rm gas}}
\def\mh{M_{\rm h}}
\def\mbh{M_{\rm BH}}
\def\mdot{\dot\mbh}
\def\mout{M_{\rm out}}
\def\mel{m_{\rm e}}
\def\mpr{m_{\rm p}}
\def\mr{{\cal R}}
\def\msink{M_{\rm sink}}
\def\msol{M_\odot}
\def\nel{n_{\rm e}}
\def\nei{n_{\rm i}}

\def\ngamma{N_{\gamma}}
\def\nscat{N_{\rm scatt}}
\def\nph{N_{\rm ph}}
\def\nt{n_{\rm T}}
\def\nub{\nu _{\rm b}}
\def\nui{\nu _{\rm i}}
\def\nut{\nu _{\rm T}}
\def\nuh{\nu _{\rm HI}}
\def\nuhe{\nu _{\rm HeI}}
\def\nuhee{\nu _{\rm HeII}}

\def\pd#1#2{\partial #1\over {\partial #2}}
\def\qe{q_{\rm e}}
\def\rch{r_{\rm ch}}

\def\rcs{r_{\rm c*}}
\def\re{R_{\rm e}}
\def\rxe{R_{\rm Xe}}
\def\rx{R_{\rm X}}
\def\rhr{\rho _{\rm h}(r)}
\def\roh{\rho _{\rm h\circ}}
\def\ros{\rho _{*\circ}}
\def\rsr{\rho _*(r)}

\def\rt{r_{\rm t}}

\def\Rsn{R_{\rm SN}}
\def\sigkn{\sigma_{\rm KN}}
\def\sigknx{\sigkn (x)}
\def\sigknnu{\sigkn (\nu)}
\def\sigt{\sigma _{\rm T}}
\def\sigi{\sigma _{\rm i}}

\def\sighc{\sigma _{\rm\circ HI}}

\def\sighec{\sigma _{\rm\circ HeI}}

\def\sigheec{\sigma _{\rm\circ HeII}}
\def\Sigx{\Sigma_{\rm X}}
\def\ss{\sigma _*}
\def\ssc{\sigma_{\rm \circ *}}

\def\tff{t_{\rm ff}}
\def\tac{t_{\rm accr}}

\def\tc{t_{\rm C}}
\def\tcc{t_{\rm cc}}
\def\tr{t_{\rm r}}
\def\td{t_{\rm d}}
\def\th{t_{\rm h}}
\def\tx{T_{\rm X}}
\def\tsx{t_{\rm x}}

\def\t15{t_{15}}
\def\vtsn{\vartheta_{\rm SN}}
\def\vx{v_{\rm X}}
\def\Zff{Z^{\rm Fe}_{\rm flow}}
\def\Zfs{Z^{\rm Fe}_{\rm\star}}

\documentstyle[11pt,aaspp4]{article}

\slugcomment{accepted version - November 17, 2000}

\lefthead{Ciotti and Ostriker}
\righthead{Cooling flows and QSOs}

\begin{document}

\title{Cooling Flows and Quasars: II.\\
       Detailed Models of Feedback Modulated Accretion Flows}

\author{Luca Ciotti\altaffilmark{1,2} and Jeremiah P. Ostriker}
\affil{Department of Astrophysical Sciences, Peyton Hall,
       08544, NJ, USA}


\altaffiltext{1}{On leave from Osservatorio Astronomico di Bologna, 
via Ranzani 1, 40127 Bologna, Italy}
\altaffiltext{2}{Also Scuola Normale Superiore, Piazza dei Cavalieri 7,
56126 Pisa, Italy}


\begin{abstract}

Most elliptical galaxies contain central black holes (BHs), and most
also contain significant amounts of hot gas capable of accreting on to
the central BH due to cooling times short compared to the Hubble
time. Why therefore do we not see AGNs at the center of {\it most}
elliptical galaxies rather than in only (at most) a few per cent of
them?  We propose here the simple idea that {\it feedback} from
accretion events heats the ambient gas, retarding subsequent infall,
in a follow up of papers by Binney \& Tabor (1995, BT95) and Ciotti \&
Ostriker (1997, CO97).  Even small amounts of accretion on a central
BH can cause the release of enough energy to reverse the central
inflow, when the Compton temperature ($\tx$) of the emitted radiation
is higher than the mean galactic gas temperature, the basic assumption
of this paper. Well observed nearby AGN (3C 273, 3C 279), having $\tx$
near $5\times 10^8$ K, amply satisfy this requirement.  In this
context, we present a new class of 1D hydrodynamical evolutionary
sequences for the gas flows in elliptical galaxies with a massive
central BH. The model galaxies are constrained to lie on the
Fundamental Plane of elliptical galaxies, and are surrounded by
variable amounts of dark matter. Two source terms operate: mass loss
from evolving stars, and a secularly declining heating by type Ia
supernovae (SNIa).  Like the previous models investigated by Ciotti et
al. (1991, CDPR) these new models can evolve up to three consecutive
evolutionary stages: the wind, outflow, and inflow phases. At this
point the presence of the BH alters dramatically the subsequent
evolution, because of the energy emitted due to the accreting gas
flow.  The effect of Compton heating and cooling, of hydrogen and
helium photoionization heating, and of
\brem recycling on the gas flow are investigated by numerical
integration of the nonstationary equations of hydrodynamics, in the
simplifying assumption of spherical symmetry, and for various values
of the accretion efficiency and supernova rates.

The resulting evolution is characterized by strong oscillations, in
which very fast and energetic bursts from the BH are followed by
longer periods during which the X--ray galaxy emission comes from the
coronal gas. For a fixed galaxy total mass and structure, the length
and the intensity of the bursts depend sensitively on the accretion
efficiency and the SNIa rate.  For {\it high efficiency} and {\it SNIa
rate} values, short and strong bursts are followed by a degassing of
the galaxy, with a consequent shut--off of the BH followed by a long
period when the mass loss from the stellar population replenishes the
galaxy, and after which a new cooling catastrophe another accretion
event takes place.  In this case high accretion rates characterize the
BH evolution, but the total mass accreted by the BH is very small. For
{\it low efficiency} and {\it SNIa rate} values, the luminosity
evolution is still characterized by strong intermittencies, but the
number of global degassing events is considerably reduced, and for
very low efficiency values it completely disappears.  The remaining
instability is then concentrated in the central galactic regions.  We
also allow for departures from spherical symmetry by examining
scenarios in which the central engine is either an ADAF or a more
conventional accretion disk that is optically thick except for a polar
region. The general property of highly unstable accretion remains
true, with central BHs growing episodically to the mass range $10^8 -
10^9\msol$ (in contrast to $\Delta\mbh\gsim 10^{10} - 10^{11}\msol$,
if feedback is ignored). In all cases the duty cycle (fraction of the
time that the system will be seen as an AGN) is quite small and in the
range $\fBH\simeq 10^{-4} - 2\times 10^{-3}$.  Thus, for any
reasonable value of the efficiency, the presence of a massive BH at
the center of a galaxy seems to be incompatible with the presence of a
long--lived cooling flow.

\end{abstract}



%

\section{Introduction}


As first revealed by {\it Einstein} observations, normal (non--cD)
elliptical galaxies, both isolated or belonging to groups and
clusters, can be powerful X--ray sources with 0.5--4.5 keV
luminosities $\lx\propto\lb^{1.8 - 2.2}$ ranging from $\sim 10^{39}$
to $\sim 10^{42}$ erg s$^{-1}$, although with a large scatter of more
than two orders of magnitude in $\lx$ at any fixed optical luminosity
$\lb\gsim 3\times 10^{10}\lsol$ (Eskridge, Fabbiano, \& Kim 1995).
From its spectral properties, this emission is associated with hot
gaseous halos around the galaxies, with mean temperatures in the range
$0.5 - 1\times 10^7$ K.

In order to explain this observational finding, the presence of
significant amounts of gas is inferred (typical values are in the
range $\mgas=10^8$--$10^{11}\msol$, see, e.g.,
\cite{form79,fjt85,cft87,fab89}),
with computed cooling times often far less than the Hubble
time. Consequently, ``cooling flow'' models have been proposed and
extensively investigated (see, e.g., Nulsen, Stewart, \& Fabian 1984;
Fabian, Nulsen, \& Canizares 1984; Sarazin \& White 1987, 1988;
Vedder, Trester, \& Canizares 1988; White \& Sarazin 1991; Brighenti
\& Mathews 1996).  While including the essential physics, these models,
however, do not provide a satisfactory account of the X--ray
properties of observed elliptical galaxies, as most of them are
observed to be much fainter in the X--rays than the models would
predict. This is specially true for homogeneous inflow models (see,
e.g., \cite{sw88}), but serious problems are also encountered by
inhomogeneous models.  The other main problem of cooling flows is that
they do not solve the question of {\it where} the cool gas is
deposited: over a Hubble time an amount of material comparable to the
mass of stars in the galactic core flows into the nucleus, while the
expected distortions of the central optical surface brightness and
velocity dispersion are not observed.  Many possible explanations for
this problem have been suggested, from local cooling instabilities
(see, e.g., Sarazin \& Ashe 1989, Bertin \& Toniazzo 1995) 
to star formation biased toward
very small masses (Mathews \& Brighenti 1999), but no one seems to be
easily acceptable. While the cooling included by the cooling flow
models is indisputable (it is observed), the question arises as to
whether heating sources within the galaxy can equal or exceed (in a
time--averaged sense) the gas radiative heat losses.

One possible solution to the previous problems was proposed by CDPR
(hereafter ``WOI'' scenario), who showed with numerical simulations of
hot gas flows in elliptical galaxies, that the heating from SNIa could
be effective in maintaining low luminosity galaxies in a wind phase
over an Hubble time (and so preventing the gas from accumulating in
the centre).  This scenario does permit one to account both for the
correlation and the data dispersion in the $\lb$--$\lx$ plane.
Nevertheless, even in this scenario the most massive galaxies develop
an inflow phase, leading to a situation similar to a standard cooling
flow, with accretion of significant amounts of gas in their center.

One of the main objections raised against the WOI scenario is the
claimed lack of detection of a sufficient amount of iron in the ISM of
some ellipticals: the SNIa are good producers of iron, and so one
would expect a strong correlation between the evolutionary scenario
for galactic gas flows and the amount of iron both in the galactic and
in the parent cluster hot gas.  For example (following the treatment
of Renzini et al. 1993), in the WOI scenario the expected present day
iron abundance in galactic flows in units of the solar iron abundance
is given by $\Zff \simeq <\Zfs>+5\vtsn$ (where $\vtsn =1$ corresponds
to the standard Tammann's rate (1982) and $<\Zfs>$ is the average iron
abundance in the stellar component).  From an observational point of
view Arimoto et al. (1997) quote $<\Zfs>\simeq 1$, while according to
Mushotzky (1999) $<\Zfs>\simeq 0.4$; moreover, Serlemitsos et
al. (1993) estimate $\Zff\simeq 0.56$ and 0.20 for NGC 1399 and NGC
4472, respectively.  As a consequence the analysis of the available
data (under the assumption of solar abundance ratios), suggests an
iron abundance in the ISM consistent with no SNIa's enrichment, and
even lower than that of the stellar component (see, e.g., Ohashi et
al. 1990, Awaki et al. 1991, Ikebe et al. 1992, Serlemitsos et
al. 1993, Loewenstein et al. 1994, Mushotzky et al. 1994, Awaki et
al. 1994, Arimoto et al. 1997, Matsumoto et al. 1997).  This is rather
puzzling: in fact, a value of $\vtsn=1/4$ is given by the current
optical estimates of the present day SNIa rate (Cappellaro et
al. 1997). However, there is an increasing evidence that more complex
multi--temperature models with a higher abundance give a better fit to
the data (Kim et al. 1996, Buote \& Fabian 1998); for example, the
latter authors find a mean value of $\Zff=0.9\pm 0.7$, while Buote
(1999) derives iron abundances up to twice solar: these figures could
be reconciled with $\vtsn=1/4$. Several evolutionary models have been
calculated with this rate (Pellegrini \& Ciotti 1998, D'Ercole \&
Ciotti 1998), resulting in the main features of the WOI scenario being
preserved, provided the amount of dark matter is properly scaled
down. In Sections 2.1 and 4.4 we will return on this point.

From another point of view, there is increasing evidence of the
presence of massive BHs at the center of most (if not all) elliptical
galaxies (see, e.g., Harms et al. 1994, van der Marel et al. 1997,
Richstone 1998, Magorrian et al. 1998).  Thus it is natural to ask
what can be expected from the interaction of a galactic gas inflow
with a massive central BH? In fact, according to the standard
interpretation of the AGN phenomenon, a massive BH is at its origin
(see, e.g., \cite{r84}).  We point out here that a strong dependence
of the {\it global} X--ray emission of early type galaxies on the
slope of their inner (optical) surface brightness profiles has been
found (Pellegrini 1999). Two recent papers considered the effect of
the gravitational field of a central BH on the density and temperature
profile of cooling flow models (Quataert \& Narayan 1999, Brighenti \&
Mathews 1999).  BT95 explored this problem assuming a release of {\it
mechanical} energy in the central regions of the galaxy, due to the
interaction between a nuclear jet produced by the accretion and the
surrounding ISM, which approach was suggested by the optically thin
status of the ISM itself, as computed by the authors with the aid of a
simple analytical model.  They indicated the circumstances under which
the central energy release might impede or reverse the infall.

As will be shown in this paper, the gas over the body of the galaxy is
really optically thin, but nevertheless the effect of {\it energy}
exchange between the emitted nuclear {\it radiation} and the gas flow
is dramatic.  This effect was studied for accretion on a BH by
Ostriker and co--workers (\cite{omccwy76,cos78}), and these authors
clearly showed how the effect of the Compton heating can be more
important than the momentum transfer itself in driving instabilities
in the accreting flow. In this context we explore in detail, by
numerical integration of the the fully non--stationary equations of
hydrodynamics, the modifications on the WOI scenario, assuming the
presence at the galaxy center of a massive black hole with initial
mass $\mbh =10^8\msol$. A summary of the main features of this new
class of models is given in CO97.

An introductory and very simple analysis of this problem can be
obtained in a semi--analytical way. For an accretion rate $\mdot$ on
the BH, the emitted luminosity can be easily estimated
\begin{equation}
\lbh = 5.67\times 10^{46}\eps\mdot\quad {\rm erg\;s}^{-1}
\end{equation}
where $\mdot$ is in $\msol\,{\rm yr}^{-1}$, and $\eps$ is the
accretion efficiency, with $10^{-3}\lsim \eps\lsim 10^{-1}$ (see
equation [9]).  It is natural to compare this number with the energy
required to {\it steadily} extract from the galactic potential well
the gas lost by the evolving stellar population at a rate $\dot\mast$:
\begin{equation}
\lgm=5.67\times 10^{40}\dot\mast\ssc^2(U_{**}+U_{\rm *h}\mr)
     \quad {\rm erg\;s}^{-1}.
\end{equation}
In equation (2) $\dot\mast$ is expressed in $\msol\,{\rm yr}^{-1}$,
the galaxy central velocity dispersion $\ssc$ is in unit of $300$ km
s$^{-1}$, and $\mr=\mh/\mast$ is the ratio between the dark matter
halo mass and the total stellar mass.  The dimensionless functions
$U_{**}$ and $U_{\rm *h}$ depend on the relative density distributions
of the galaxy stellar component and of the dark matter halo, and are
of the of order of unity.

The ratio between these two luminosities is then given by:
\begin{equation}
{\lbh\over\lgm}\simeq 6.7\times 10^{16}
                      {\eps\mdot\over (1+\mr)\lb\ssc^2\t15^{-1.3}}
               \simeq {10^6\eps\over (1+\mr)\ssc^2},
\end{equation}
where $\t15$ is the age in 15 Gyr units, and the stellar mass losses
have been expressed as function of the present galaxy blue luminosity
(in $\lsol$), $\dot\mast\propto\lb\t15^{-1.3}$ (see Section 2.1 for a
detailed description). Numerical simulations shows that a normal
elliptical in the inflow phase can accrete at a rate $\sim 1\msol$
yr$^{-1}$ or higher (accretion rates of tens or hundreds solar masses
per year are common when inflow are established at early times in
giant ellipticals, see, e.g., CDPR). Assuming this value as an
estimate for $\mdot$, it is immediatly realized that the energy
emitted by accretion can easily be {\it orders of magnitude larger}
than that formally required to interrupt the cooling flow, even
assuming very low efficiency values.  Note that this simple energetic
argument predicts a greater effect of $\lbh$ on the gas flows hosted
by small galaxies with respect to those on giant ellipticals; a
decrease of the dark matter content goes in the same sense.  From the
first formula at the r.h.s. of the equation above it is also clear
that, at early times, {\it more} energy is required to extract the gas
from the galaxy potential well, but this is compensated by the same
fact that more gas is available.  If we assume (erroneously) a
stationary situation, where the produced $\dot\mast$ is accreted on
the center, then the second expression in the equation above is
obtained, again showing the importance of the radiation emitted in the
galactic energy budget.

Obviously, the true situation is by far more complicated.  In fact,
three main questions arise about the reliability of equation (3). The
first concerns how much of $\lbh$ is actually {\it trapped} by the
galactic gas and so effectively available for its heating: it is clear
indeed that for extremely low opacities, the {\it effective} energetic
balance would be unable to produce a {\it global degassing}, although
strong instabilities may occur. For the moment we will take as a minimal
estimate of the absorption that due to the inverse Compton effect,
inelastic scattering of the ISM electrons interacting with the high
energy photons emitted by the central BH.

We estimate the expected gas opacity using a density profile
representative of the inflow models computed in CDPR,
$n(r)=n_0/(1+r^2/r_{\rm gas}^2)^{b/2}$.
Defining $\Delta\lbh$ to be the fraction of the accretion luminosity
emitted at the galaxy center that interacts with the
galactic hot gaseous halo, from equation (A10) we obtain
\begin{equation}
{\Delta\lbh\over\lbh}\simeq
{2\sqrt{\pi}\,\Gamma [(b-1)/2]\over\Gamma (b/2)}
{\nel\over\nt}{\cx\kb\tx\over\mel c^2}
\sigt n_0 r_{\rm gas}\simeq 3\times 10^{-14}\tx n_0,
\end{equation}
having for simplicity assumed a gas temperature $T$ much lower than
the temperature $\tx$ associated with the spectral distribution of
$\lbh$ (see equation [10]), $r_{\rm gas}=300$ pc, and $b=2$.  For
example, for $n_0=10^2$ cm$^{-3}$ (a common density value in the
central regions of cooling flows, see, e.g., CDPR), and $\tx=10^9$ K,
we find a total absorption of $3\times 10^{-3}$, in accordance with
that given by BT95.  So, combining equations (3) and (4), we see that
we can expect major degassing or strong intermittencies in the
accretion (and in the coronal $\lx$ evolution as well) when
\begin{equation}
\tx\gsim 5\times 10^{-4}{\lb\ssc^2 (1+\mr)\over n_0\eps\mdot\t15^{1.3}}
\approx 3.3\times 10^7{\ssc^2 (1+\mr)\over n_0\eps}\quad {\rm K}.
\end{equation}
One can easily imagine situations so that the r.h.s.  of equation
above is as low as $10^6$ K or as high as $10^9$ K. This shows at once
that the emitted radiation is effective in producing a strong feedback
on the accretion flow, but that numerical simulations are needed in
order to understand the flow evolution in specific cases.  Note here
that, for a given value of $\mdot$ the absorption of energy in the
ambient gas is proportional to {\it both} the efficiency, $\eps$, and
the Compton temperature $\tx$. Thus, so long as $\tx >> T_{\rm gas}$,
{\it the basic assumption of this paper}, all that matters is the
product $\eps\tx$ and we will be free, in the work that follows to
imagine $\tx$ reduced by a factor of ten and $\eps$ increased by the
same factor. We will test this scaling in Section 4.2.4.

A more complex situation may however arise, if the accretion happens
onto an accretion disk, and so it may be that much of the emitted
radiation is funneled into a small solid angle, so that it is not seen
by the inflowing gas, i.e., the assumption of spherical symmetry is
qualitatively in error. As a consequence, only a fraction of the
surrounding galactic hot gas can see the emitted radiation.  We have
to remember, hovewer, that the nearly isotropic {\it Thomson}
scattering on the electrons can back--heat the galactic gas.  One can
estimate the order of magnitude of this effect reducing $\Delta\lbh$
as given in equation (4) by a fiducial factor $1-\exp (-\tau)\simeq
\tau$, where $\tau=\sigt\int_0^{\infty}n_{\rm e}(r)dr$
and only the back--scattered radiation is permitted to heat the ISM.
Allowing for this further reduction in the efficiency of the Compton
pre--heating on the gas flows and using the same figures as above for
the gas distribution, one finds $\tau=\pi\sigt n_0 r_{\rm gas}/4\simeq
0.05$, and so the lower temperature limit increases by a factor of 20.
A reasonable interval of $\tx$ seems to be still available to
produce at least strong intermittencies in the gas flow.

The second problem with a naive use of equation (3) is the fact that
the balance between the cooling and the heating of the gas in the
central galactic regions can be determinant for the flow evolution.
In fact it is important to stress that $\Delta\lbh/\lgm>1$ is only
necessary for a global galaxy degassing: if the gas is able to radiate
efficiently the absorbed energy (perhaps as low energy photons), then
it is plausible that only a more or less strong perturbation of the
flow in the inner galactic regions will result as a consequence of the
accretion.  However, a simple estimate of the radiating efficiency of
galactic gas flows can be obtained directly from observations. The
brightest observed X--ray galaxies are characterized by luminosities
$\lx\lsim 10^{42}$erg s$^{-1}$, and we take this as an upper limit for
the gas flows radiating efficiency. This is well below $\lbh$
predicted by equation (1), and so the cooling time is longer than the
heating time during the most intense accretion events.

The third problem with equation (3) is that it gives the order of
magnitude of the {\it steady} power necessary to expell the gas.  On
the contrary, it is known that Compton heating can produce strong
intermittencies on accretion flows: what is the effect of a strong but
{\it impulsive} heating on galactic inflows?  Only numerical
simulations can possibly answer these questions.  In any case, from
this preliminary and very simple analysis, we can expect that the
presence of a massive BH at the galaxy centre will be very significant
in modifying the gas flow.

The paper is organized as follows. In Section 2 we describe in detail
the models and the input physics, in Section 3 we briefly describe the
equations of the problem and the numerical code used for their
integration.  In Section 4 we describe the numerical results for
various choices of the parameters, corresponding to various physical
assumptions.  Finally in Section 5 we discuss the results and their
implications.

\section{The Model}

\subsection{The Galaxy Model}

The density distribution of the galaxy models have the same profile as
in CDPR, in order to maintains continuity with this previous work and
so appreciate better the effect of introducing the feedback of the
central BH on the gas flows. A detailed description of how their
parameters are fixed in order to reproduce galaxies following the
Fundamental Plane relation is given in CDPR.  The stars are
distributed following a truncated King profile (\cite{k72})
\begin{equation}
\rsr ={\ros\over(1+\eta^2)^{3/2}},
\end{equation}
where $\ros$ is the central stellar density, $\rcs$ is the galactic
core radius, and $\eta\equiv r/\rcs$.  This distribution is truncated
at the tidal radius $\rt=\delta\rcs$ to give a finite galaxy mass. In
all the explored models we fix $\delta=180$. The dark matter halo is
described by a quasi--isothermal distribution,
\begin{equation}
\rhr ={\roh\over 1+\eta ^2/\beta ^2},
\end{equation}
where $\beta\equiv\rch/\rcs$, $\roh$ is the central dark matter density
and the tidal radius $\rt$ is assumed to be the same as for the
stellar distribution.  We define two other dimensionless parameters,
namely the ratio between the central densities $\gamma\equiv\roh/\ros$,
and the ratio between the total masses $\mr\equiv\mh/\mast$.

The stellar mass loss rate and the SNIa rate are the main evolving
ingredients of these models, and again, for an extensive discussion,
we refer to CDPR. Here it is sufficient to recall that in the code the
stellar mass losses -- the source of {\it fuel} for the activity of
the BH -- follow the exact prescriptions of the stellar population
synthesys, but for quick calculations the approximation
${\dot\mast}(t)\simeq 1.5\times 10^{-11}\lb\t15^{-1.3}\;\msol$
yr$^{-1}$ can be used, where as before $\lb$ is the present galaxy
blue luminosity in $\lsol$. The quantity entering the hydrodynamical
calculations is not directly $\dot\mast$, but the specific rate of
mass return $\als\equiv\dot\mast /\mast$. The SNIa rate is
parameterized as $\Rsn (t)=0.88\times 10^{-12} h^2\vtsn\lb\t15^{-s}$
yr$^{-1}$, where $h\equiv H_{\circ}/100$ km s$^{-1}$ Mpc$^{-1}$.  When
$\vtsn=1$ and $\t15=1$, the standard SNIa Tammann's rate is recovered
(\cite{tam82}); we use $h=1/2$. Assuming for each supernova event an
energy release in the ISM of $10^{51}$ erg s$^{-1}$, the energy input
per unit time over all the galaxy body is then given by
\begin{equation}
\lsn (t)=7.1\times 10^{30}\vtsn\lb
          \t15^{-s}\quad\quad {\rm erg}\,{\rm s}^{-1}.
\end{equation}
Besides energy, supernovae provide also mass. We assume that each SNIa
ejects $1.4\msol$ of material in the ISM, and so the total specific
rate of mass return becomes $\alpha=\als +\asn$, with $\asn (t) =
{1.4\,\Rsn (t)/\mast}$. Finally, each SNIa produces $0.7\msol$ of
iron.  Observationally, the ICM contains $0.01-0.02\msol$ of iron for
every $\lsol$ of the galaxy (star) population of the cluster, with no
appreciable trend with cluster richness: in the parmeterization we
adopted for the SNIa rate this can be obtained with $\vtsn=1/4$ and
$s\simeq 1.6-1.7$ (see, e.g., Renzini et al. 1993, and references
therein).

\subsection{The Central BH}

At the onset of the cooling catastrophe a flow of gas is suddenly
accreted by the BH, and this produces a burst of energy from the
galaxy center. The {\it instantaneous} bolometric luminosity
associated with the accretion is:
\begin{equation}
\lbh(t) =\eps \,\mdot(t)\,c^2,
\end{equation}
where $\mdot$ -- the accretion rate on the BH -- is assumed to be {\it
positive} if the flow is accreting, and $\eps$ is the accretion
efficiency, usually spanning the range $0.001\lsim\eps\lsim 0.1$. A
fundamental ingredient in the treatment of the interaction between the
galactic gas flow and the radiation $\lbh$ is its (normalized)
frequency distribution given by $\fznu\equiv\tilde\fz (x)/\nut$, where
$x\equiv\nu /\nut$ and $\nut\equiv\mel c^2/h$ is the Thomson
frequency.  We adopt
\begin{equation}
\tilde\fz (x)={\xi_2\over\pi}\sin\left[{\pi(1-\xi_1)\over\xi_2}\right ]
\left ({\nub\over\nut}\right )^{\xi_1+\xi_2-1}{x^{-\xi_1}\over
(\nub/\nut)^{\xi_2}+x^{\xi_2}},
\end{equation}
where $h\nub$=1 MeV is the spectrum break energy, and $\xi_1=0.5$ and
$\xi_2=0.7$; these values are consistent with the observed spectral
indices of X--ray emission spectra of QSOs (see, e.g., Lu \& Yu 1999).
From the adopted spectral distribution given by equation (10) the
Compton temperature is $\tx\simeq 10^9$ K (see Appendix A). As noted
earlier, we can approximately scale results obtained with this $\tx$
to a lower value with corresponding increase in $\eps$.  Of course,
the scaling argument breaks down when $\tx$ is lower than the mean gas
temperature, and our model cannot be applied: in this case the
galactic gas will experience Compton cooling, and this case is briefly
discussed in Section 5.  Here we present observational evidence
supporting our choice of high $\tx$ associated with the spectra of
some well known AGNs in ellipticals. For example, observations of 3C
273 in the range 150-800 keV by OSSE (McNaron-Brown et al. 1997), show
a double slope frequency distribution, with $\xi_1=0.62$,
$\xi_2=0.9^{+1.6}_{-0.6}$, and a break energy of $300\pm 100$
keV. These values correspond (with a break energy of 200 keV) to
$\tx\simeq 9.3\times 10^8$ K, $\tx\simeq 5\times 10^8$ K, and
$\tx\simeq 10^8$ K, for $\xi_2=0.4$, $\xi_2=0.9$, $\xi_2=2.5$,
respectively.  However, the above computation does not include the
effect of the so--called ``UV bump'' (a common feature in the AGNs
spectra), and so a more accurate estimate of $\tx$ is obtained by
using the spectral distribution over the whole observed energy
range. We did this computation for 3C 273, possibly the best observed
AGN, with available data over the energy range $6.1\times 10^{-8}$ eV
-- $1.0\times 10^9$ eV (T\"urler et al. 1999), and we obtained
$\tx\simeq 3.2\times 10^8$ K. We note here that when limiting the
integration to data below 10 keV (200 keV) we obtain $\tx\simeq
1.2\times 10^6$ K ($\tx\simeq 5.3\times 10^7$ K). The same computation
for 3C 279 (Maraschi et al. 1994) gives $\tx\simeq 6.6\times 10^8$ K
(whole spectrum, $1.6\times 10^{-5}$ eV -- $9.9\times 10^9$ eV),
$\tx\simeq 5\times 10^7$ K (200 keV cut--off), and $\tx\simeq 10^6$ K
(10 keV cut--off). Finally, for Mrk 421 during quiescence (Macomb et
al. 1995), we obtain $\tx\simeq 10^9$ K (whole spectrum, $1.98\times
10^{-5}$ eV -- $2.1\times 10^{12}$ eV), $\tx\simeq 2\times 10^8$ K
(200 keV cut--off), and $\tx\simeq 3\times 10^7$ K (10 keV cut--off).
Note the good agreement of the above results with the $\tx$ estimate
obtained from the X--ray background.  Madau \& Efstathiou (1999) found
that the average photon energy of the X--ray cosmic background is 31.8
keV, which translates to a $\tx\simeq 3\times 10^8$ K: then, if one
assumes that the typical redshift of emission was $z = 1\div 2$, there
is another factor of two or three bringing one to the range $\tx\simeq
6\times 10^8 - 10^9$ K. This is interesting, because there is
increasing evidence that the origin of the X--ray background (at
energies larger than 1 keV) is primarily due to AGN type sources
(Mushotzky et al. 2000, Giacconi et al. 2000).

The main effect of the emitted radiation on the galactic gas flows
considered in our simulations is the Compton heating (and cooling) of
the gas: this is due directly to the central radiation source and
indirectly to the recycling of the \brem radiation produced by the gas
heated to very high temperatures by the BH activity. We consider also
the effect of photoionization in changing the cooling function at low
temperatures, and finally the heating of cold gas due to the
photoionization of hydrogen and helium.  At each radius the opacity of
the gas to the Compton heating is calculated, using for the electrons
the Klein--Nishina cross--section.  Obviously, photons do not exchange
only energy with the gas, but also momentum, and this leads to the
introduction in the momentum equation of hydrodynamics of an {\it
effective gravitational field}, that for $\lbh$ greater than the
Eddington luminosity can revert the velocity of the gas flow. The
treatment used to describe numerically all these physical phenomena is
given in Appendix A.

\subsection{A Numerical Estimate of $\lbh$}

A non trivial problem is represented by the numerical treatment of
$\lbh (t)$, a function of the physical quantities that are necessarily
defined only at finite distance from the galaxy center, far from the
Schwarzschild radius of the BH.  A physically correct boundary
condition would be to have some grid points inside the radius $\rx$
where the potential energy of the gas (due to the presence of the BH)
equals the thermal energy of the gas associated with the Compton
characteristic temperature $\tx$. A simple calculation shows that
\begin{equation}
\rx={2G\mpr\mbh\over 3\kb\tx}\simeq
    3.5\times 10^{-10}{\mbh\over\msol}\quad {\rm pc},
\end{equation}
a very small scale length when compared to the galactic size; moreover
the associated time scale is $\tsx =\rx /\vx$,
where $\vx^2=G\mbh/\rx$, and so
\begin{equation}
\tsx=9.7\times 10^{-8}\,\mbh \left ({\tx\over 10^9 {\rm K}}\right)^{-3/2}
\quad {\rm yr}
\end{equation}
of the order of ten years for a for $\mbh =10^8\msol$.  Clearly these
figures prevent the possibility of simulating the gas flow from inside
$\rx$ to hundreds of kpc over time scales of Gyrs.  The problem
becomes even worse during the flaring activity: numerical models, in
which we calculated the evolution of the gas flow in a region
extending up to few parsecs from the BH, and in which 10 grid points
are contained inside $\rx$, frequently show that the Courant time and
the heating time become of the order of a {\it day}, unequivocally
forbidding any attempt of a global numerical simulation extending to
all the characteristic temporal and spatial scales. In a subsequent
paper we investigate at very high (spatial and temporal) resolution
the gas flow close to the BH, i.e., we give a complementary analysis
of the problem investigated in this paper, on time scales of the order
of years and less. Here we are interested in the behavior of the gas
flow on galactic scales and over times of Gyrs, and we believe we have
captured with our description the macroscopic properties of the
evolution.

We circumvent the problem of obtaining in a consistent way $\lbh (t)$
from the hydrodynamical quantities at the first grid point exploring
two limiting cases for the description of the accretion. In the first
case, that we call {\it instantaneous} accretion, after a time delay
equal to the free--fall time from the first grid point, the inflowing
gas is accreted from the BH and disappears in the central sink. For
the adopted galaxy model $\tff$ can be easily computed for $r<\rcs$:
\begin{equation}
\tff(r)=\sqrt{3\over 4\pi G \ros(1+\gamma)}
        \int _0^1 {\sqrt x\;dx\over\sqrt {(1-x)({\cal D} + x + x^2)}},
\end{equation}
where ${\cal D}=3\mbh/2\pi\ros (1+\gamma)r^3$.  The integral can be
evaluated explicitly in terms of elliptic functions, and equals $\pi
/2$ for ${\cal D}=0$ and $\pi/3$ for ${\cal D}=1$. For ${\cal
D}\to\infty$ the integral is asymptotic to $\pi/2\sqrt{{\cal D}}$.

In the second case, which we call {\it smoothed} accretion, we mimic
the presence of an accretion disk around the BH, where the fuel is
stored during the inflow phases and subsequently released on the
BH. In our spherically symmetric models, this complex phenomenon is
simply simulated introducing the {\it accretion time}, $\tac$, in
which the gas from the accretion disk disappears in the BH.  With this
assumption a given amount of gas that crosses the first grid point at
the time $t$ is not accreted instantaneously at $t+\tff$ but with an
exponential decay of characteristic time $\tac$.  For example, in the
$\alpha$--disks, $\tac =O(t_{\rm orb}/\alpha)$, where $t_{\rm orb}$ is
the orbital time at that radius (Shakura \& Sunyaev 1973).  We use the
parametrization $\tac=\kappa\tsx$, and the method used for the
numerical computation of $\lbh (t)$ in this case is described in
Appendix B.

Obviously for a fixed efficiency and for a given quantity of accreted
mass the {\it energy} released in the two scenarios is the same, and
so the {\it luminosity} produced by the smoothed accretion is lower
but emitted over a longer time (a few $\tac$).  What may be the effect
of a smoothed accretion on a short time--scale is not hard to imagine:
in fact, it is natural to expect a {\it reduced} effect on the gas
flow with respect to the instantaneous accretion model with the same
efficiency $\eps$. In other words, we expect something similar to a
reduction of the {\it effective} efficiency.  Over longer time--scales
a guess on the gas behavior is not so simple: from the reduction of
$\lbh$ one would expect the flow to become more similar to that with
instantaneous accretion and smaller $\eps$, i.e., a more uniform
flaring activity possibly without a major galactic degassing. But it
is also true that after few $\tac$, the energy emitted is the same as
in the case of the instantaneous accretion. This smoother energy
release over a longer time could produce a {\it stronger} tendency to
global degassing. Only numerical simulations can answer which of these
two competing effects will finally dominate.

\section{The Hydrodynamical Equations}

As in CDPR, the evolution of the galactic gas flows is obtained
integrating the time--dependent Eulerian equations of hydrodynamics
with source terms:
\begin{equation}
{\pd \rho t}+\nabla \cdot (\rho v)=\alpha\rho_*,
\end{equation}
\begin{equation}
{\pd m t}+\nabla \cdot (mv)=-(\gamma-1)\nabla E+\gef\rho,
\end{equation}
\begin{equation}
{\pd Et}+\nabla \cdot (Ev)=-(\gamma-1)\,E\nabla \cdot v-\lrad+\lh+
\alpha\rho_*({\cal E}_{\circ} +{1\over 2}v^2),
\end{equation}
where $\rho$, $m$, and $E$ are respectively the gas mass, momentum and
internal energy density, and $v$ is the gas velocity.  The ratio of
the specific heats is $\gamma =5/3$, and $\gef (r)$ is the effective
acceleration experienced by the gas, considering the effect of the
radiation field (see Appendix A).  ${\cal E}_{\circ}= 3\kb
T_{\circ}/2\mu\mpr$ is the injection energy per unit mass, where
$T_{\circ}(r,t)=[\asn (t) T_{\rm SN}+\als (t)T_*(r)]/\alpha(t)$ is the
gas injection temperature, that describes the heating of the gas by
the thermalization of the SNIa blast waves and of the stellar mass
losses due to their relative motion with respect of the ISM. $T_{\rm
SN}$ is obtained as in CDPR, while the injection temperature due to
the stellar random motions $T_*(r)\equiv\mu\mpr\ss^2(r)/\kb$ is
increased near the galaxy center with respect to the models discussed
in CDPR, due to the the presence of the BH. The radial behavior of the
stellar velocity dispersion can be explicitly found for the King
density distribution by solving the Jeans equation:
\begin{equation}
\rsr\ss^2(r,\mbh)=\rsr\ss^2(r,\mbh=0)+
{G\mbh\ros\over\rcs}\left[{1+2\eta^2\over\eta\sqrt{1+\eta^2}}-
{1+2\delta^2\over\delta\sqrt{1+\delta^2}}\right],
\end{equation}
where for simplicity we assumed global isotropy of the velocity
dispersion tensor.  The first term on the r.h.s. of equation (13) is
the velocity dispersion without the BH (see, e.g., \cite{c93}).  In
the energy equation $\lrad$ is the cooling function, modified to allow
for the ionizing effects of $\lbh$ on the cold gas, and the total
heating function is $\lh=L_{\rm C}+L_{\rm ph}+\lbr$.  $L_{\rm C}$ is
the Compton heating (or cooling), $L_{\rm ph}$ describes the effect of
the photoionization heating on hydrogen and helium, and finally
$L_{\rm Br}$ describes the effect of recycling the \brem radiation
emitted from the inner part of the galaxy on the external part of the
gas flow.  A detailed discussion of the heating and cooling functions
is given in Appendix A.

The numerical engine of the code is essentially the same as in CDPR,
i.e., a first order, eulerian up--wind scheme (\cite{vanl77}), on
which several gestures towards second order accuracy are implemented.
In the computation of galaxy models the grid has 120 zones and extends
up to 500 kpc. In order to obtain an acceptable spatial resolution in
the inner regions a geometrical progression for the mesh size is
adopted. At the outer boundary the usual outflow conditions are
assumed. The first active grid point for the discussed models is
placed at 20 pc from the galaxy center, and this, at variance with
what was done by BT95, is an {\it active} grid point, in the sense
that the fluid velocity is self--determined by the flow and by the
heating during the flare activity, and not imposed as a boundary
condition. Some models were run with higher resolution, i.e., with the
first active grid placed at 10, 5, or 2 pc from the center, obtaining
a good qualitative agreement of the resulting evolution with that
obtained using the larger grid.

The sink of the hydrodynamical quantities due to the BH is simulated
by removing inside the first grid point the equivalent quantity of
density, momentum, and energy that disappears during the accretion
events. We assume the model galaxies to be initially devoid of gas,
i.e., $\rho (r)=0$ for $t=0$.  This is certainly a simplification of
the real physical situation in the early evolutionary stages of
elliptical galaxies, when intense star formation takes place. In this
way only the gas originated by stellar mass loss is actually
considered.  Our initial condition, $\rho (r)=0$, is therefore
equivalent to assuming that galactic winds are established by SN
activity (mainly type II), early in the evolution of elliptical
galaxies. The choice of the time--step is determined by five
characteristic time--scales related to different aspects of the
problem: the Courant time $\tc$, the cooling time $\tr=E/\lrad$, the
density source time $\td=\rho/|\dot\rho|$, the heating time
$\th=E/\lh$, and finally the free--fall time from the first grid
point, $\tff$.  The numerical time--step is taken to be some specified
fraction (0.5 or less) of the minimum of the previous times computed
over the integration grid, and the equations are integrated using the
time--splitting method.


\section{The Results}

In this Section we present the results of the simulation of the gas
flow for a model similar to model KRM discussed in CDPR.  For this
reference model we adopt $\lb=5\times 10^{10}\lsol$ and $\ssc=280$ km
s$^{-1}$, obtaining from the FP relation $\re=4.2$ kpc (and so
$\rcs=350$ pc, $\ros=7.15\times 10^{-19}$ g cm$^{-3}$, and
$\mast=2.8\times 10^{11}\msol$, see CDPR).  The dark matter halo is
characterized by $\mr=7.8$ and $\gamma=1.25\times 10^{-2}$ from which
$\beta=4.2$. The BH accretion efficiency is assumed to be $\eps=0.1$,
and for the SNIa parameters we adopt $(\vtsn ,s)=(2/3,1.5)$, a choice
for which initially there is ample SN heating to sustain a supersonic
wind. The accretion mechanism adopted for this model is instantaneous
accretion, and the free--fall time from the first grid point (20 pc)
is $\tff\simeq 1.5\times 10^5$ yr.  In Table 1 all the model
parameters, together with the derived quantities, are reported (model
\#1). Above it, we show the properties of model \#0, which is
identical to model \#1, except that all feedback is turned off; the
assumed efficiency is $\eps=0$.  We note here that the expected
present day iron abundance in the gas flow of this model would be
$\simeq 4.3$ times the iron solar abundance, considerably larger than
the current observational estimates. However, we present here this
model in order to have a direct comparison with the same model
discussed in CDPR; moreover, in Sections 4.2.3 and 4.4 we present
other models with a reduced SNIa rate.

In the following Subsections we will explore the changes in the
characteristics of the evolution by decreasing the efficiency $\eps$
from the reference value (0.1) to 0.01 and 0.001, by increasing the
dark--halo mass to reduce the time $\tcc$ of the global cooling
catastrophe or decreasing the SNIa rate, and finally by mimicking in a
qualitative way the effect of the presence of an ADAF solution or
optically thick accretion disk in the galaxy center.

\subsection{The Evolution of Model \#1}

A common characteristic of all models discussed in this paper is
obviously their similarity with WOI models up to $\tcc$, a substantial
central activity starting only after the onset of the cooling
catastrophe. For this reason here we do not describe in detail the
flow evolution before $\tcc$, which the interested reader can find in
CDPR, where the three main phases of {\it wind}, {\it outflow}, and
{\it inflow} are exaustively discussed.  In this paper we concentrate
instead on the evolutive phases for $t>\tcc$, discussing the previous
evolution only where necessary to understand the new results
presented. We parenthetically note that the wind and outflow phases
are required if one is to understand the high metal content of gas in
cluster of galaxies (see also Section 4.4).

In Fig.1ab the time evolution of the mass budget of model \#1 is
shown, sampled at time intervals of 1 Myr.  More specifically, in
Fig.1a the dotted line represents the mass of gas, $\mgas (t)$ in
solar masses contained inside the galaxy truncation radius
$\rt=\delta\rcs =63$ kpc, and the solid line the mass accreted on the
BH, $\msink$. In Fig.1b, the {\it rates} of variation of various
masses are shown, all in solar masses per year: the dotted line
represents the mass return from the stellar population $\dot\mast$,
the dashed line the rate of mass loss at the tidal radius
$\dot\mout=4\pi\rt^2\rho(\rt)v(\rt)$, and the solid lines is the
accretion rate on the BH, $\mdot$. The temporal evolution of these
quantities is easily understood. During the early {\it wind} phase the
flow is characterized by supersonic outward velocities driven by SNIa
heating, and more gas is lost at $\rt$ than produced inside the galaxy
by the stellar population. The wind phase ends due to the decreasing
of the specific heating, that smoothly leads to the {\it outflow}
phase, i.e., to a subsonic degassing. However, a time comes when the
outflow velocity has decreased so much that the internal gas density
begins to rise, in spite of the continuing decrease in $\dot\mast$.
Correspondingly, the total amount of gas inside the galaxy reaches a
minimum precisely when the rate of mass loss from the galaxy drops
below the rate of injection of fresh gas $\dot\mast$. In model \#1
this minimum is reached at $\sim 5$ Gyr, as can be clearly seen in
Fig.1a.

\placefigure{fig1}

In Fig.1c the flow evolution is seen from an energetic point of view:
the short--dashed and long--dashed lines are respectively the power
supplied to the gas by the SNIa explosions ($\lsn$, equation [8]), and
the power required to {\it steadily} extract from the galaxy potential
well the gas lost by the evolving stellar population ($\lgm$, equation
[2]).  The solid line represents the coronal X--ray luminosity of the
gas, $\lx$, that would be observed in the 0.5--4.5 keV range.  Note
how up to $\tcc$ the time evolution of $\lx$ follows strictly that of
the gas mass inside the galaxy. Superimposed is the luminosity
produced by the accretion, as given by equation (1).

Returning to a discussion of the gas flow, we see that, as more and
more gas builds up inside the galaxy, its central density rises, the
radiative losses become increasingly more important, a smaller
fraction of the SNIa heating is left to drive the flow, and in turn
the flow deceleration causes a faster increase of the gas
density. Inevitably, a moment comes when, at the center, the heating
rate drops below the cooling rate, and a central cooling catastrophe
suddenly reverses the direction of the flow in the inner region of the
galaxy: the transition to an inflow regime is virtually
instantaneous. In the models described in CDPR this phase was called
{\it inflow}, and after this point the evolution of their models is
smooth, as can be seen in Fig.2, where $\lx (t)$ of a model identical
to model \#1 but without the central BH is shown (dotted line, model
\#0 in Table 1) superimposed to that of model \#1 (solid line).  The
value of the time averaged X--ray luminosity $<\lx >$ is slightly
below $10^{41}$ erg s$^{-1}$, comparable with typical observed X--ray
bright ellipticals (see Table 2). In comparison, model \#0 (without
feedback) has an average luminosity of $10^{42}$ erg s$^{-1}$,
considerably above observed sources.

\placefigure{fig2}

For the specific choice of parameters made for model \#1, the cooling
catastrophe happens at $\tcc\simeq 9$ Gyr, when negative infall
velocities appear near the center, and rapidly exceed $\sim 100$ km
s$^{-1}$ at a few hundreds pc away from it. From this time onwards the
flow evolution is substantially altered by the presence of the BH with
respect to the models discussed in CDPR.  Immediately after $\tcc$ an
accretion rate of $\mdot\gsim 100\;\msol$yr$^{-1}$ is established, as
can be seen from Fig.1b (and its time expansion in Fig.3a).  The
accretion on the BH produces a strong energetic feedback, the gas in
the central regions of the galaxy is heated by the emitted radiation
approximately to the Compton temperature $\tx$, and as a consequence
it starts to expand and its density decreases by orders of magnitude.
The effect of this strong heating near the BH is a correspondingly
strong reduction of the gas radiative losses, and the accretion on the
BH suddenly stops: only after an episode of local cooling can the next
accretion event take place, and the cycle re--starts. When the amount
of accreted gas is sufficient, a strong shock wave finally reaches the
galactic tidal radius, reducing dramatically the amount of the total
gas present inside the galaxy.  Only after the time required for the
stellar mass losses to replace the gas lost, can the gas density
increase sufficiently for the radiative losses to become again
important over galactic scales and produce a new cooling catastrophe.
During brief periods the central BH emits in the range $10^{46}\gsim
\lx /{\rm erg}\,{\rm s}^{-1}\gsim 10^{45}$, compatible with typical AGN
luminosities.

The described evolution can be followed from a local point of view in
Fig.1d, where the evolution of the gas (number) density (dotted line)
and temperature (solid line) at the inner active grid point (20 pc) is
shown. All the discussed phases can be easily recognized: particularly
note how $\dot\mout$ soon after each accretion event overcomes
$\dot\mast$, so decreasing the total amount of gas inside the galaxy.
From Fig.1a one can also note how the total amount of mass which has
been accumulated by the BH after 15 Gyr is relatively low and
compatible with typical luminous AGN, only $\sim 10^8\msol$: on the
contrary, the same model without the central BH (model \#0), whose
$\lx$ is represented in Fig.2 with the dotted line, at $t=15$ Gyr had
accumulated in its central regions more than $\simeq 1.5\times
10^{10}\msol$ of gas, more than observed in any purported central BH.

As in the case of the evolutionary phases preceeding the cooling
catastrophe, namely the wind and the outflow phases, also during the
AGN--like phases the coronal X--ray emission of the gaseous halo
follows the evolution of the mass of the gas inside the galaxy, as can
be seen in Fig.1c, where the solid line is $\lx$. The dotted line in
the same figure is $\lbh$, and it is interesting to observe how $\lx$
and $\lbh$ are anti--correlated: during the accretion phases the total
luminosity is completely dominated by the nuclearly concentrated
$\lbh$, on the contrary during the quiescent BH phases the galaxy
emission is due only to the hot gas $\lx$, and so it is diffuse as
observed in X--ray elliptical galaxies.  As a consequence, the
luminosity evolution of model \#1 is composed by two spatially and
temporally complementary phases, characterizing the two different flow
regimes: one, very short, in which the total luminosity is dominated
by the central activity, and another, much longer, dominated by the
coronal gas diffuse luminosity.  Observationally, we would identify
one phase with AGNs and the other with nascent cooling flows.

The temporal evolution of $\lx$ is shown in Fig.2 (solid line) where
one see also its large excursion (even more than a factor of 100) from
immediately after the nuclear bursts compared with before the
accretion events. A very important parameter characterizing the
different phases of the galaxy luminosity evolution is the duty cycle
$\fBH$ of the central engine, that we define as\footnote{Note that for
a burst of time duration $\Delta t_b$ and constant luminosity,
$\fBH=\Delta t_b/(15-\tcc)$; it is the fraction of time that the BH is
``on''.  For a luminosity profile $L(t)\propto A-\cos (\omega t)$ and
$A\geq 1$, $f=2A^2/(1+2A^2).$}
\begin{equation}
\fBH\equiv {(\int_{\tcc}^{15}\lbh\,dt)^2\over
        (15-\tcc)\int_{\tcc}^{15}\lbh^2\,dt}.
\end{equation}
For model \#1, we found $\fBH\simeq 10^{-4}$; thus this model would be
seen as an AGN for only a very small fraction of the Hubble time.

A final comment on the global $\lx$ evolution is in order to stress
how an intrinsic mechanism able to produce a large scatter in the
diffuse X--ray luminosity of ellipticals is naturally available when a
strongly discontinuous nuclear activity is present: statistically the
X--ray observations can catch $\lx$ in any point of its variation, and
this variation, as can be clearly seen in Fig.2, can be as high as the
variation during the wind and outflow phases. A discussion of this
point is postponed in Section 5.

We move now to describe the small--scale temporal evolution of the
accretion events.  In fact, as can be seen from Fig.1, each burst
shows a rich temporal sub--structure that is interesting to discuss in
greater detail.

In Fig.3abc we plot the temporal expansions of the corresponding
panels in Fig.1bcd, sampling the represented quantities at time
intervals of $10^4$ yr.  In Fig.3a the instantaneous accretion rate is
shown, and its sub--structure is apparent: these short--time
intermittencies, of increasing intensity, are due to reflected shock
waves in the inner regions of the galaxy that carry fresh gas to the
BH. After many of these precursors a larger amount of gas is finally
accreted, and the galaxy can effectively lose a significative fraction
of its gaseous halo, with ejection velocities at $\rt$ of the order of
100 km s$^{-1}$ or more. The relative role of cooling and heating is
apparent in Fig.3c.

\placefigure{fig3}

A very small but representative sample of the radial profiles of the
hydrodynamical quantities at various important evolutionary phases is
shown in Fig.4. In the right column the gas flow is approaching
$\tcc$: each line is separated by the successive (in the following
order: solid, dotted, short-dashed, long-dashed, dotted-dashed) by 10
Myrs, starting at 9.01 Gyrs. Thedetailed description of this phase
(the end of the outlfow phase in WOI models) is given in CDPR.

The three panels of the central column cover the evolution of the flow
in the time interval 9.06-9.10 Gyrs, i.e., during a period of strong
instabilities (as can be seen from Fig.3c).  Note how large (i.e.,
supersonic) positive and negative velocities are present well inside
the effective radius, and how cold shells carry new fuel to the
central BH: one of such events can be seen following the cold shell
represented by the dotted line, and after that the short dashed line
showing a low density cavity expanding at high velocity.  Soon after
another dense shell forms and falls on the center (dotted-dashed
line). Sometimes, outward moving blast waves catch up to previous
blast waves, and reflect back.

Finally, in the last column, the time sequence starts after a
considerable accretion event, at 9.2 Gyr (see Fig.3c). The emitted
energy is able to ``clean'' the central galactic regions: the
perturbations propagate towards the galactic tidal radius, and the
density profile returns smooth (and characterized by a well defined,
centrally flat, low density profile). 

\placefigure{fig4}

As already pointed out in the Introduction, we allowed in our
simulations for the effects of the gas opacity and photoionization
heating.  For model \#1 it turns out that the fraction of $\lbh$
trapped by the hot gas around the galaxy never exceeds $10^{-2}$, and
many times reaches values as low as $10^{-4}$, confirming the
estimates obtained in the Introduction and by BT95.  We conclude that
the gas is always optically thin, but nevertheless we performed all
simulations using the correct treatment for the opacity.  Moreover,
the same model with and without photoionization heating (as well as
{\it all} models we explored) produced essentially the same evolution.
We collect in Table 1 the relevant overall properties of model
\#1, to be compared with variant models. Comparison between 
model \#0 and the others shown makes clear that a feedback effective
efficiency even as low as $\eps\tau=10^{-3}\times 10^{-3}=10^{-6}$, is
enough to radically change the outcome from the zero feedback case.

\subsection{Exploring the Parameter Space}

We move now to comment on some models in which the basic parameters
(efficiency, dark matter amount, SNIa rate, Compton temperature) are
changed with respect to model \#1. Actually, we computed many more
models than those discussed in the following, but those presented can
be considered a representative set of the various kinds of evolution
we found.

\subsubsection{Reducing the Efficiency}

In this Section we describe the effects on model \#1 gas flow by
assuming now $\eps=0.01$ and $\eps=0.001$ (labelled models \#2 and
\#3).  These two models are also explored in the scenario of
istantanous accretion.  The evolution for $t <\tcc$ is the same
already discussed, and so we start our description directly at
$t=\tcc$, expecting that a reduction in the efficiency will lead to a
flow evolution in which the probability of major degassing events is
reduced if not impossible.  As discussed in the Introduction this does
not also mean that the central, short--time bursts disappear: on the
contrary, it is possible that now a {\it greater} fraction of the flow
evolutionary time is spent at high $\lbh$, with the consequent
increase of $\fBH$.

For $\eps=0.01$ (model \#2), the model evolution is qualitatively
similar to model \#1, as can be seen from the value of the various
quantities given in Table 1.  Note that the total mass accreted by the
BH has increased now to $3.1\times 10^8\msol$, due to the lower
efficiency that produces a weaker feedback on the gas flow for the
same mass accretion rate.

\placefigure{fig5}

The same trend is also shown by model \#3, where $\eps=0.001$: in
fact, at 15 Gyr the BH mass is increased by $\sim 4.3\times
10^8\msol$, and more bursts are present.  For what concerns the BH
duty cycle, we found $\fBH =2.7\times 10^{-4}$ and $\fBH =3.8\times
10^{-4}$, for model \#2 and \#3, respectively. However,
notwithstanding these differences, the temporal structure of the
luminosity evolution of these three models remains qualitatively the
same, as can be seen comparing the three panels in Fig.5, where the
energetic of models \#1,2,3 is shown.  Even for these two latter
models the total fraction of trapped $\lbh$ in the hot gas remains
very low during all the evolution, with values in the range
$10^{-4}-10^{-3}$.  Relevant overall properties of models
\#2 and \#3 are displayed in Table 1. Especially to be noted is the 
relatively small range in the mean gas X--ray luminosity $<\lx>$.  As
the efficiency is reduced by two orders of magnitude $<\lx>$ increases
from $7.6\times 10^{40}$ erg s$^{-1}$ to $1.5\times 10^{41}$ erg
s$^{-1}$, all values being far below the no feedback value of $\simeq
10^{42}$ erg s$^{-1}$ (model \#0, see Table 2).

\placetable{tbl-1}

\subsubsection{Early Cooling Catastrophe: 
               Increasing the Dark Matter Content}

In this Section we describe the results obtained for the evolution of
three models correspondingly identical to models \#1,2,3 in their
stellar amount and accretion efficiencies ($\eps=0.1$ in model
\#4, $\eps=0.01$ in model \#5, and $\eps=0.001$ in model \#6), but in
which the dark matter to visible matter mass ratio is increased from
7.8 to 10, in order to obtain an earlier cooling catastrophe. As a
consequence, $\tcc$ decreases from $\sim 9$ Gyr to $\sim 4.2$ Gyr.

\placefigure{fig6}

In the three models, the luminosity evolution is of the same type of
that presented by model \#1, but the central energy bursts are much
more frequent (see Fig.6).  This is a direct consequence of the fact
that now the galaxy potential well is deeper with respect to that of
model \#1, as qualitatively shown by equation (3); moreover, as in
models with reduced efficiency described in Section 4.2.1, the
probability of major degassing events is reduced if not impossible.
Thus, the total mass accreted by the BH in these models is higher than
in the corresponding cases with less massive dark haloes: in fact, we
found values of $5.8\times 10^8\msol$, $7.0\times 10^8\msol$, and
$6.9\times 10^8$, for models \#4,5,6, respectively. Note however that
a model without feedback and identical in all other properties to
models \#4,5,6, accretes on its central BH $\Delta\mbh =2.5\times
10^{10}\msol$ over an Hubble time.  The same comment applies to the
duty cycle, for which we found $\fBH =2.1\times 10^{-4}$, $2.6\times
10^{-4}$, and $2.5\times 10^{-4}$, for models \#4,5,6, respectively.
It is interesting to note that the model presenting the strongest
tendency towards global degassing is {\it not} model \#4 (the model
with the higher efficiency), but model \#5 (with $\eps=0.01$).  This
is also the model that accreates more mass, and with the higher
duty--cycle. Relevant overall properties of these models are displayed
in Table 1.

\subsubsection{Early Cooling Catastrophe: Reducing the SNIa Rate}

As anticipated in the Introduction, one of the main objections raised
against the WOI scenario is the claimed detection of low iron
abundance in the ISM of ellipticals. Prompted by this fact, we also
present three models (models \#7,8,9) in all respects identical to
models \#1,2,3, except for the fact that now $\vtsn =1/4$ instead of
2/3 (see equation [8]). Moreover, we also computed three more models
identical to models \#7,8,9, but with the ratio between the dark
matter halo mass to stellar mass reduced from 7.8 to 3 (models
\#10,11,12).  Others models analogous to models presented in this
Section, but with a smaller $\lb$ and central velocity dispersion, are
described in Section 4.4.

\placetable{tbl-2}

With the reduction in the SNIa heating, the cooling catastrophe
happens now at $\tcc=0.25$ Gyr. For the three explored efficiencies,
the effects on the temporal behavior of the bursts are dramatic: the
separation between the long phases of quiet evolution and the short
bursts with rich temporal sub--structures chracterizing model \#1 is
practically disappeared, and only a very strong BH activity is
present, with some intervals during which a more global degassing is
approached.  Very few and short outgassing events are still present,
but the time between successive major accretions is substantially
reduced.  This is shown in the upper panel of Fig.7, where the
luminosities are plotted in the representative time interval from 9
Gyr to 9.5 Gyr, for model \#7 ($\eps=0.1$): because of this the
qualitative aspect of the evolution of the flow remains of this kind
over all the evolution for the three models.

\placefigure{fig7}

As stressed above, the temporal sub--structure of the accretion is
maintained, i.e., the accretion is still locally unstable.  This fact
is rich in consequences, the most important of which is the increase
of duty cycle, with respect to models with ``late'' cooling
catastrophe: in fact, now $\fBH =1.5\times 10^{-3}$, $1.8\times
10^{-3}$, and $1.6\times 10^{-3}$ for models \#7,8,9.  Moreover, the
total mass accreted by the BH in these three models is $3.5\times
10^9\msol$, $4.9\times 10^9\msol$, and $5\times 10^9$, respectively,
substantially higher than in the models previously presented.  For
comparison, we computed also the mass accreted on the central BH by an
identical model with $\eps=0$, and we found $\Delta\mbh \simeq
2.5\times 10^{11}\msol$, even higher (as expected) than our first
computed no feedback model \#0.

The behavior of models \#10,11,12 is very similar, and their relevant
properties are listed in Table 1; again, for the corresponding model
with $\eps=0$, it turns out that $\Delta\mbh\simeq 1.1\times
10^{11}\msol$. From the values reported in Table 1 one can see how the
decrease of the ratio $\mh/\mast$ (maintaining all other model
parameters fixed) produces a reduction of the total mass accreted by
the BH (cfr. also models \#1,2,3 with \#4,5,6).  As illustration, the
luminosity evolution of model \#10 is shown in Fig.7 (middle panel).

\subsubsection{Reducing T$_{\rm X}$}

As anticipated in the Introduction, a simple argument suggest that
models with constant $\eps\tx$ should behave in the same way, so long
as $\tx >> T_{\rm gas}$. In order to test quantitatively this
prediction, we computed a few models with the Compton temperature
reduced by a factor of 10: models \#1r and \#2r are derived from
models \#1,2 assuming $\tx=10^8$ K, and in model \#1rr $\tx=5\times
10^7$ K. The resulting evolution is very similar to that of previous
models, so that even substantial reductions of the Compton temperature
(as long as $\tx$ is higher than the mean gas temperature) produce a
strongly unstable evolution.  The overall properties of these models
are presented in Table 1.  The first apparent result is that, as
expected, the central BH of each ``temperature reduced'' model
accretes more mass than the corresponding ``hotter'' model. However,
the scaling is not perfect. This is due to the complex nature of
accretion. In fact, as discussed in Section 4.1, each accretion event
is made of two different dynamical phases. In the first, the accretion
is determined by the cooling of gas, and the scaling argument
perfectly applies. The second phase, (where a substantial fraction of
the gas is actually accreted) is characterized by many accretion
events (with a substantially lower accretion rates) caused by
reflected shock waves propagating in the central region of the
galaxy. After the first, cooling dominated accretion event, the gas
temperature in the central regions of the galaxy can reach values as
high as $10^9$ K, and so the scaling argument no longer applies
rigorously.  For example, computing the {\it mean} accretion gas
temperature at the first grid point (20 pc) for a complete accretion
event in model
\#1r we found
\begin{equation}
{\tx\over <T_{\rm gas}>}\equiv
{\int\mdot \tx/T_{\rm gas}\;dt\over \int\mdot\;dt}\simeq 1.9.
\end{equation}
Moreover, approximately 2 per cent of the total mass accreted by the
central BH in the accretion event is at temperature {\it higher} than
$\tx$, i.e., in Compton cooling regime.  This explain the semingly
counterintuive result that the total mass accreted by these models
with reduced $\tx$ {\it decreases} with decreasing efficiency, at
variance with all the other models: obviously the effect of Compton
cooling is stronger for the higher efficiency models.  In any case,
the main features of all the other models are retained: the time spent
by the galaxy in a AGN state is still very small, the accreted mass is
low, and the time averaged $<\lx>$ is substantially lower than that of
model \#0 (see Table 2).

\subsection{Non--Spherical Accretion: ADAF and 
            Optically Thick Accretion Disks}

The assumption of spherical symmetry adopted in our models is
certainly not realistic for an accurate modelling of the accretion
phenomenon. This is specially true near the BH, where, in the case of
a non--zero angular momentum, the gas can be stored in an accretion
disk, and then smoothly released with some characteristic time scale
on the BH. Another interesting possibility recently suggested is the
so--called {\it Advection Dominated Accretion Flow} (see, e.g., NY95),
where the resulting efficiency can be very low, due to nature of the
solution (radiation from relatively low density gas, see Park \&
Ostriker 1999, and references therein).  Certainly the answer to this
problem can be obtained only using at least a 2D numerical code (in
the interior parts of the flow), but despite this obvious limitation,
we believe that some very simple numerical experiments can be of some
real interest. In fact, given the relatively low specific angular
momentum assumed in the ADAF solutions, they would be quite spherical
in the domain explored by our simulations ($r\geq 20$ pc). Thus our
computations plausibly represent the feeding (outer boundary
condition) for the ADAF models.

\subsubsection{ADAF}

We start by presenting the results of a simulation in all ways similar
to model \#1, but where we replaced the central engine with the mildly
unspherical ADAF solution.  Following NY95, we modify $\lbh$ as
\begin{equation}
\lbh(t) ={\dot m\over\alpha^2+\dot m}\eps\mdot(t)c^2,
\end{equation}
where we use $\alpha=0.3$ and $\dot m\equiv \eps c^2\mdot/\ledd$.  At
low accretion rates the effective efficiency is proportional to
$\eps\dot m$ in the ADAF solution as it is in the equivalent spherical
case (Shapiro 1973, Park \& Ostriker 1998) but it limits out to $\eps$
for high $\dot m$.  Model \#13 is analogous to model \#1, i.e.,
$\eps=0.1$, as assumed by NY95.  The resulting evolution of this model
is characterized, quite surprisingly, by a {\it more} effective
degassing than that showed by model \#1. In fact in this case, when
the accretion starts, the central BH is {\it less} reactive than in
the analogous model \#1, due to the effect of the modulating factor in
the equation above. As discussed in Section 2.3, this allows a {\it
larger} quantity of gas to be accreted in any single burst, and as a
consequence a stronger effect on the galactic gas flows (see Fig.8,
upper panel).  The total mass accreted by this model is $2.1\times
10^8\msol$, and its duty cycle is $\fBH=1.2\times 10^{-4}$.

Two more models with the central ADAF, but with the efficiency reduced
to $\eps=0.01$ (model \#14) and $\eps=0.001$ (model \#15), show a
reduced tendency towards global degassing. Their duty cycle is now
$\fBH =1.3\times 10^{-3}$ for model \#12 and $\fBH =3.3\times 10^{-3}$ for
model \#13, and, correspondingly, the total mass accreted by the BH is
$4.4\times 10^9\msol$ and $2\times 10^{10}\msol$. Their luminosity
evolution is shown in Fig.8 (middle and lower panel, respectively).
Again, the reduction of $\eps$ produces dramatic consequence on the
model evolution, but the flow is still strongly unstable.

Thus, while we stress again the fact that the models here presented
can give at best a qualitative understanding of the gas evolution in
galaxies, we note that there are compelling evidences that ADAF
solutions, {\it at least when residing in galactic cooling flows}, are
unstable to Compton pre--heating. Since the duty cycle is invariably
very low, the ADAF (``AGN'') flow would be seen only a small fraction
of the time, perhaps providing a viable solution to the problem of the
apparent lack of QSO activity in most ellipticals.

\placefigure{fig8}

\subsubsection{Optically Thick Accretion Disks}

Finally we turn to a substantially different case.  To look at a
maximally non spherical inner solution we consider the possible
presence of physically thin, optically thick accretion disk.  Two
effects are considered here: the first is the fact that now a
substantial fraction of the emitted luminosity $\lbh$ can be shielded
by the disk and thus emitted in only a narrow cone emanating from the
central BH, and, second, the modulation in the accretion due to the
storage of the gas in the accretion disk will alter the temporal
behavior.  As discussed in the Introduction, we model the effect of
the disk on the gas heating computing at each time step the Thomson
opacity $\tau$ of the galactic ISM, and then allowing only the
back--scattered fraction, $1-\exp(-\tau)$, of $\lbh$ to be thermalized.
As described in Section 2.3, the second effect is characterized by the
introduction of the parameter $\tac=\kappa\tsx$, which prescribes the
rate of the fuel release from the disk to the BH. The detailed
description of how this release is modeled in the code is given in
Appendix B. In these models $\eps=0.1$, as is commonly assumed in
accretion disks.

The two models here presented are structurally identical to model \#1,
and are evolved adopting $\tac=10^2\tsx\simeq 10^3$ yr, and
$\tac=10^4\tsx\simeq 10^5$ yr (models \#16 and \#17, respectively).
The resulting evolution was quite insensitive to the exact value of
$\tac$, but very much affected by the reduction of the effective
Compton heating.  In fact, altough strong variability is present (see
Fig.7, lower panel, where the luminosity evolution of model \#17 is
shown), no major degassing events are visible.  In both cases the duty
cycle, the fractional time during which a quasar would be seen in the
center remains small, precisely $\fBH\simeq 5.3\times 10^{-2}$ and
$5.4\times 10^{-2}$ for models \#16 and \#17, respectively, but
substantially higher than all others duty cycles listed in Table 1. In
any case, also for these models, the diffuse gas remains optically
thin over all the simulation. The mean X--ray luminosity of the gas in
the two models is nearly identical, i.e., $<\lx >\simeq 3\times
10^{41}$ erg s$^{-1}$, and the final BH mass is $\simeq 1.6\times
10^9\msol$.

So, with this small number of models designed to investigate the
possible effect of non--spherical accretion, we reached the quite
counterintuitive conclusion that, albeit characterized by high
efficiency, the optically thick accretion disks are {\it less}
effective than ADAF solutions in expelling the gas from the galaxy
potential well. However in both cases the essential results of the
spherical case are retained.  Only a small fraction of the ambient ISM
is accreted to the center with, for the bulk of the gas, the
time--averaged Compton heating balancing the radiative cooling, so
that the classic {\it global} cooling flow catastrophe does not occur.

\subsection{Changing Galaxy Optical Luminosity and Central Velocity 
Dispersion}

All models presented in the previous Sections are characterized by the
same optical luminosity and central velocity dispersion. How does the
gas flow evolve in galaxies with different optical luminosity and
central velocity dispersion? In particular, is the evolution of models
characterized by $\vtsn=2/3$ peculiar to high SNIa rate models or is
the WOI scenario still valid if we reduce the present day SNIa rate to
the current (optically) estimated $\vtsn =1/4$ (and rescaling
accordingly the amount and/or distribution of the dark matter
content)? We present here few selected models of an extensive work (in
progress) aimed at investigating the properties of a large set of
galaxy models. In all of the following models $s=1.6$ in order to
match the observed iron in the ICM (see Renzini et al. 1993). As
already anticipated by CDPR, the main result is an extreme
sensitiveness to the value of the central velocity dispersion. For
example, model \#18 ($\lb=3\times 10^{10}\lsol$, $\ssc=200$ km
s$^{-1}$) remains in the {\it wind} phase over all its
evolution. Model \#19 has the same parameters as model \#18, but a
slightly higher central velocity dispersion (i.e., a slightly deeper
potential well), $\ssc=210$ km s$^{-1}$. In this case $\tcc\simeq 9.5$
Gyrs, while in model \#20 ($\ssc=220$ km s$^{-1}$), $\tcc\simeq 4.6$
Gyrs. In model \#21 ($\lb=4\times 10^{10}\lsol$, $\ssc=210$ km
s$^{-1}$) $\tcc\simeq 9.6$ Gyrs, while in the model \#22 ($\ssc=220$
km s$^{-1}$) $\tcc\simeq 7.2$ Gyrs.  The evolution of these models is
very similar to that of the models presented in the previous Sections:
for example, in Fig.9 (upper panel) the evolution of the X--ray
luminosity of model \#21 is shown, and the similarity with that of
models \#1,2,3 is apparent.  Also the duty cycles span the same range
of values: $\fBH =(4.7,\;5.6,\;1.7,\;2.7)\times 10^{-4}$ in models
\#$(19,\;20,\;21,\;22)$

Of course, due to the dependence of the stellar mass loss on the
parent galaxy optical luminosity, the accreted mass in the galaxy
center is correspondingly reduced, with $\Delta\mbh\simeq
(0,\;6.5\times 10^7,\;2.9\times 10^8)\msol$ in models
\#(18,19,20), and $\Delta\mbh\simeq (4.8\times 10^7,\;1.2\times 10^8)\msol$ 
in models \#(21,22). Note the dependence of $\Delta\mbh$ on $\ssc$
(see Section 5).

\placetable{tbl-3}

In all the models presented above $\tx=10^9$ K, $\eps=0.1$, and
$\mr=2$. A reduction of $\tx$ obviously increases the mass accreted in
the center, but does not alter dramatically the evolution (as far as
its value is higher than the mean gas temperature, the basic
assumption of this paper): for example, in model \#21r (Fig.9, lower
panel) we assumed $\tx=2\times 10^8$ and we obtained $\Delta\mbh\simeq
1.4\times 10^8$ and $\fBH=6.2\times 10^{-4}$. Relevant overall
properties of these models are displayed in Table 3.

\section{Discussion and Conclusions}

In this paper we have investigated the behavior of the galactic gas
flows in early type galaxies, assuming at their center the presence of
a massive BH growing with the accretion of matter and affecting the
inflow through feedback. The interaction of the radiation produced by
accretion with the hot, X--ray emitting gas of the halo has been
studied, considering the effects of Compton heating and cooling,
\brem recycling, variable ionization state in the cooling function and 
hydrogen and helium photoionization heating. The basic assumption of
these models is that the Compton temperature of the radiation emitted
by the accreting material is substantially higher than the mean gas
temperature of the galactic ISM, consistent with the finding that
nearby AGN in elliptical galaxies have $\tx\simeq 5\times 10^8$ K.

The characteristics of the flow and of the resulting luminosity are
studied by changing the efficiency of conversion of the accreted mass
into radiation, the Compton temperature of the emitted radiation, the
rate of SNIa explosions and the amount of dark matter.  A qualitative
exploration is also performed in order to understand the effect of the
presence of a physically thin, optically thick accretion disk, where
the inflowing gas is stored before being released with a
characteristic time scale on the BH and radiation is emitted only in a
cone, or alternatively an ADAF solution, characterized by
accretion--dependent low luminosity.

The resulting evolution of the flow -- and consequently of the emitted
luminosities, both coronal ($\lx$) and central ($\lbh$) -- is
sensitive to all these parameters, but {\it the gas flows are found
invariably unstable to Compton heating}. Before discussing the
variants let us note the properties in common to all of the models:

\begin{itemize}

\item At late epochs (comparable to the current Hubble time), all models
show a type of relaxation oscillation. They spend a small fraction of
the time ($10^{-2} - 10^{-4}$) in an AGN state of high BH luminosity
($10^{45} - 10^{47}$ erg s$^{-1}$), and the bulk of the time with the
AGN ``turned off'' and the ambient gas emitting X--rays in what would
appear to be a proto--cooling phase state.

\item Mass dropout and accreted mass are small compared to naive estimates
based on gas mass/cooling time because the time averaged heat input
from the central BH balances the gas radiative cooling.

\item The central BH grows by episodic accretion up to 
a mass in the observed range ($10^{8.5} - 10^{9.5}\msol$) in all giant
ellipticals, even though only $\approx 10^{-3}$ of them will be
observed in an AGN phase.

\item Due to the self regulating nature of the feedback, all the computed 
observables are relatively insensitive to the assumed efficiency (in
the range investigated).

\end{itemize}

For all the computed models, of which only a small but representative
sample is described in this paper, the gas opacity to $\lbh$ remains
very small in all evolutionary phases, as predicted by BT95 and by our
introductory analysis: the fraction of the emitted luminosity
effectively absorbed by the halo gas is of the order of
$10^{-4}-10^{-3}$, and never larger than $10^{-2}$. Moreover, the
effect of photoionization heating is found to be negligible, due to
the high temperatures reached by the gas during the central bursts.

For high efficiency values ($\eps\simeq 0.1)$, the flow evolves
through strong oscillatory phases: short bursts of accretion that last
few Myrs, with a rich temporal sub--structure, and in which the
radiation emitted by the central source overcomes the luminosity
emitted by the coronal gas, are followed by longer periods (of the
order of Gyrs) in which the BH is quiescent, and the X--ray luminosity
is due only to the coronal gas. The quiescent phases are a consequence
of the galaxy degassing following the strongest central accretion
events, when a substantial fraction of the galaxy hot gas is lost.  We
note that the high metallicity seen in the ICM of typical rich
clusters provides strong observational evidence for significant wind
and outgassing phases.  The decreased gas density reduces dramatically
the radiative losses, and the inflow stops. After the time required
for the stellar mass loss to replenish the galaxy and so to restore
the radiative losses, a new global cooling catastrophe takes place,
the BH re--ignites, and a cycle restarts. The maximum luminosities
reached during the accretion events can be of the order of
$10^{47}-10^{48}$ erg s$^{-1}$, but the duration of the single
accretion events are extremely short. Correspondingly, the accretion
rates can reach values as high as $10^2\msol$ yr$^{-1}$, but the total
mass accreted by the BH at the end of the simulation is a few
$10^8\msol$, very low compared to the amount that the galaxy expels
during the BH bursts. $\lx$ is strongly correlated with the amount of
hot gas inside the galaxy, and so after a degassing event the galaxy
X--ray luminosity can drop by a factor of 100 or more, and
successively increase together with the gas amount inside the galaxy.
The luminosity of the central AGN and the ambient gas are strongly
anticorrelated.

Reducing the efficiency, the time elapsed in the phases of BH activity
becomes longer and longer, and for very low efficiency values the
global degassing events completely disappear, leaving a continuous
flickering of $\lbh$. The values of $\lbh$ in these circumstances are
still high, $\sim 10^{46}$ erg s$^{-1}$ at peaks, and the final mass
of the BH is much higher than in case of high efficiency, reaching
values as high as $10^9\msol$ or more.

Models in which the SNIa rate was decreased or the dark matter halo
mass was increased, and so $\tcc$ is anticipated at early times, have
been successively investigated. The differences between high and low
efficiency accretion hold also for this class of models, and are
essentially the same as for late times cooling catastrophe. The only
quantitative differences concern the time elapsed between two
successive global degassing events. In this case the time is shorter,
consistently with the faster time evolution of the mass return from
the stellar galactic population at early times. For these models the
central BH grows to over $10^{11}\msol$ in case of no feedback.
However, it was shown that the epoch of the cooling catastrophe is
extremely sensitive to the value of the model central velocity
dispersion, and that even galaxies with a SNIa rate consistent with
the current optical estimates and of considerable optical luminosity
($\lb\geq 3\times 10^{10}\lsol$) can easily develop a {\it late}
cooling catastrophe and successive global degassing events.

Two more classes of models were explored in semi--quantitative
fashion, in order to explore the main effects of non--spherically
symmetric accretion.  In the first class, we studied the effects of
the presence of a physically thin, but optically thick accretion disk
around the BH, where the infalling gas is stored and successively
released with some time scale. The same disk funnels most of the
radiation into a polar cone from which only the Thomson
back--scattered component is able to Compton heat the gas. Now, with
the overall efficiency reduced by a factor $\tau\sim 10^{-2} -
10^{-1}$, the solution is similar to that of low efficiency models.
In these models, the major degassing events are suppressed, and only a
(strong) flaring activity persists. The central BH can accrete up to
nearly $10^{10}\msol$ over the Hubble time.  More interesting is the
case of a central ADAF.  In fact, at variance with simple
expectations, the accretion--dependent efficiency characterizing these
solutions produces an efficient galaxy degassing.

As an important point, we point out here that all the simulations
confirm the simple energetic expectations worked out in the
Introduction, i.e, the fact that the energetic balance in our model is
critical: in fact, we found both models with global degassing and
models presenting only strong instabilities in the accretion, but
without major degassing events.

A delicate issue which applies to all computed models concerns the
temporal structure of the sub--bursts.  We sampled $\lbh$ and $\lx$ at
very high temporal resolution, and we found that each sub--burst is
temporally extended for several time--steps.  This fact is interesting
for two reasons: the first is that the sub--bursts are not numerical
artifacts, produced by the time--discretization used by the numerical
scheme of integration, but real phenomena of the models. The second
fact is that each sub--burst extends temporally approximately over the
free--fall time from the first active grid--point (Section 3.1). So,
not only is this phenomenon well described numerically, but also {\it
physically}, at least consistently with the adopted spatial
discretization.  Obviously, by reducing the grid spacing in the inner
regions the temporal resolution can be increased. We performed many
simulations with a reduced grid spacing, and the final result was that
we obtained a better temporal resolution of each burst, finding that
each is subdivided into an increasing number of shorter
sub--bursts. At the same time, the total time of each burst, the
accreated mass, the emitted luminosity are in good agreement with
those derived using the standard grid, and so we believe we have
captured the physics of the phenomenon.  In any case, it is clear that
only using a grid with extremely high spatial resolution, that samples
the flow well inside $\rx$, can the substructure of a burst be studied
in full detail. This kind of simulation, with several grid points
inside $\rx$ and the outermost one placed at few pc, will be presented
in a later work. It represents the complementary exploration to that
performed in this paper, allowing a comparison with observations of
AGN variability.

How can our results be interpreted in the more global context of the
X--ray luminosity evolution of the bulk of elliptical galaxies?  In
the modeling of gas flows {\it without} a central engine, two
different possibilities have been explored in the past: in the first
the SNIa heating is so low that the galaxies are not able to expel
their coronal gas, and so all galaxies should host an inflow. The
three main problems encountered by this scenario are that the expected
X--ray luminosity is substantially higher than that observed, the
related difficulty in explaining the observed scatter of $\lx$ at
fixed optical luminosity, and finally the disposal of the gas that
cools and accumulates in the galaxy centre in a standard cooling flow.

These problems could be partially solved in the WOI scenario, in which
a time decreasing SNIa rate is sufficient to sustain a wind. In this
class of models the observed scatter in the $\lb$--$\lx$ plane is a
direct consequence of the deepening of the galactic potential well
with increasing $\lb$, as dictated by the Fundamental Plane. This
scenario solves also the problem of the mass accreted in the galaxy
center over an Hubble time by cooling flow galaxies, being the major
fraction of the gas lost by the stars ejected from the galaxies in
their early wind phase, when also the mass return is higher.  However,
as discussed in the Introduction, objections have been raised on the
validity of this scenario, on the ground of the apparent lack of
detection of a sufficient amount of iron in the ISM of some
ellipticals. Obviously, for a null SNIa rate, the WOI scenario does
not apply, and the cooling flow solution, together with its unsolved
problems, would remain the only workable model.  In any case, even
with enhanced supernova activity up to the current optical estimates,
the cooling catastrophe is only delayed, not avoided, so the mass
disposal problem remains for massive galaxies.

Following the observational consequences that can be derived from the
class of models presented in this paper, a solution to the previous
problems can be obtained quite easily, mainly due to the strong and
temporally discontinuous variability of $\lbh(t)$ and $\lx(t)$.

The first observational consequence of our models is that the duty
cycle of the central engine, even in models unable to produce
substantial global degassing events, is very low, never exceeding
$10^{-1}$ and typically of order $10^{-2}$ to $10^{-3}$.  For high
efficiency accretions this quantity can be as low as $10^{-4}$, and
its value increases for low efficiency.  The details for each model
are given in Table 1.  These numbers are small, and could be an
alternative explanation to that advocated by Fabian \& Rees (1995) in
order to explain why the nuclei of elliptical galaxies are normally
not luminous sources as a result of the accretion of the hot gas by
the central BH.

A second interesting aspect of the presented models is that the total
mass accreted on the galactic center over an Hubble time, even in the
case of an early cooling catastrophe, is low compared to the same
quantity for a pure cooling flow solution (or even for inflow solution
in the CDPR models, see Table 1). We note that the accreted masses are
comparable (with the exceptions of models \#0 and 15), to the masses
observed in massive central black holes, so the {\it feedback
modulated accretion flows} could provide a natural way to grow black
holes to the observed masses (see, e.g., Kormendy \& Ho 2000).

Recently it has been determined that $\mbh$ correlates extremely well
with $\ssc$, with $\mbh\propto\ssc^{4\div 5}$ (see, e.g., Ferrarese \&
Merritt 2000, Gebhardt et al. 2000): we note here that $\Delta\mbh$ in
galactic models differing only in the value of their central velocity
dispersion (models \#19,20 and \#21,22 in Section 4.4) also increases
with $\ssc$. Of course, the scatter of $\Delta\mbh$ of models in Table
1 is larger than that observed: a reduced scatter would be predicted
if (for example) the accretion efficiency and the present SNIa rate
were fixed numbers.

\placefigure{fig9}

A third aspect to be mentioned is the evolution of the hot gas X--ray
surface brightness profiles: in fact one of the many problems of the
cooling flow models is that the resulting X--ray profiles $\Sigx (R)$
are cuspy, at variance with the observations (\cite{sa89}).  As
already pointed out by BT95 and CO97, in this new class of solutions
characterized by feedback from the galactic central regions, the
profiles have a well defined core, as in the case of the outflow
solution of CDPR, except during the central burst phases, when the
$\Sigx$ shape presents many features.  In Fig.10 we show the
surface brightness profiles of a model representative of the case in
which global degassing events take place (model \#1, upper panel), and
of a model presenting only strong temporal variations of the accretion
(model \#7, lower panel).  The profiles are sampled at ten different
times, equally spaced of 0.5 Gyr, starting from 9.5 Gyr (after the
cooling catastrophe of model \#1). The heavy lines in both panels are
the X--ray surface brightness profile of the corresponding models
without the central BH, at a representative time (12 Gyr).  The
profiles are obtained using a projection routine based on the
Raymond--Smith thin plasma code (Raymond \& Smith 1977). The absence
of the central cusp in these models is apparent, except for the very
cuspy profile in the lower panel, representing a model caught during
an accretion event.

The total time in which $\Sigx$ of model \#1 is significantly
disturbed is much less than 10 per cent of the time after the cooling
catastrope ($\sim 6$ Gyr), considerably more in model \#7.  Thus the
duty cycle of Compton perturbations in $\Sigx$ is much greater than
that of the central $\lbh$.

The presence of these large--scale features is clearly an important
observational test for the scenario explored in this paper. Certainly
models such as \#7 (e.g., \#8,9,10,11,12, lower panel in Fig. 10) are
characterized by very noisy $\Sigx$ profiles, and so observations
probably are already in conflict with them. On the contrary, the
profiles of models similar to \#1 (e.g., \#2,3,4,5,6,13,21,1r,2r,21r
and others, upper panel in Fig. 10) are considerably smoother,
essentially over all the galaxy lifetime. For these latter models the
only significant feature in the $\Sigx$ profile is the short--lived,
X--ray bright shell visible in the higher profile in Fig. 10 (upper
panel). The observational properties of this feature have been
described in CO97.

\placefigure{fig10}

A fourth important point is the aid these models can give to the
solution of the observed scatter in the $\lb$--$\lx$ plane, as
preliminarly discussed in CO97.  In fact, in our models $\lx$ shows
large variations in its intensity (see, e.g., Fig.2, and Table
1). However, the definitive answer to the problem of the $\lb$--$\lx$
scatter could be more complicated. In fact, recent observational
findings (Mathews \& Brighenti 1998, Fukugita \& Peebles 1999), albeit
based on just 11 galaxies and all of relatively high X--ray
luminosity, seem to show a correlation $\lx\propto\rxe^{\alpha}$,
where $\rxe$ is the X--ray effective radius and $\alpha$ is a positive
number ($\alpha =0.6\pm 0.3$ in Mathews \& Brighenti 1998).  As well
known, this is at odds with the predictions of cooling flows and CDPR
inflow models: only during CDPR {\it outflowing} models this
correlation could be established. Also the X--ray luminosity of the
models here presented is (statistically) anticorrelated with $\rxe$,
as can be clearly seen in the upper panel of Fig.10, where the lower
profiles (i.e., low $\lx$) are also the more extended (i.e., large
$\rxe$).  A possible explanation for $\lx-\rxe$ relation could involve
environmental effects, as proposed by Mathews \& Brighenti.


Summarizing, if the WOI scenario applies, small galaxies do not
experience in their life any cooling catastrophe, and their luminosity
evolution would be the same as described in CDPR, with low $\lx$ and
no significant nuclear activity.  The nuclear activity should
predominate in medium-high luminosity galaxies, but the probability of
catching a peak of $\lbh$ is small, of the order of a few per cent or
less, as obtained from the performed numerical simulations with high
efficiency; a larger fraction of the galaxies {\it hosting a BH at
their center} should be caught in a nuclear outburst if the efficiency
of the accretion is low.  Note that, if the origin of the QSO
phenomenon is associated with the initial cooling catastrophe, then we
have a model for the ``clock'': for example, in a standard
$\Lambda$CDM universe ($h=0.65$, $\Omega_{\rm m}=0.3$,
$\Omega_{\Lambda}=0.7$, see, e.g., Bahcall et al. 1999) $\tcc
=(4,\,7,\,9)$ Gyr corresponds to $z=(1.7,\,0.8,\,0.5)$.

Certainly the presented models have some weaknesses: first of all the
assumption of spherical symmetry overall, but we think that the
present investigation is sufficiently promising that a more demanding
effort is justified. In the real accretion phenomena non axisymmetric
effects in the infalling gas are very likely to occur outside of the
central regions. Also, outflows may occur in the polar directions
(cfr. BT95 and Park \& Ostriker 1999) as jets, thus changing the total
amount of accreted mass with respect to the spherically symmetric
estimates here presented. The interaction between these jets and the
hot gaseous halo can be investigated only with a 2D code.  Another
important question that requires (at least) a 2D treatment is the
possibility of instabilities in the gas flow.  Particularly, from our
simulations it appears that a transient cold shell of material
surrounds a hot central bubble during the flaring activity, especially
in the case of low efficiency.  This would evolve into a
Rayleigh--Taylor instability for the shell, that certainly will break,
permitting cold fingers of material to accrete on the BH.  It is then
clear that a fully 2D hydrodynamical analysis is required for a better
understanding of this accretion problem. The unstable nature of the
gas flows in elliptical (without central BHs) has been found by 2D
numerical simulations, (D'Ercole \& Ciotti 1998, D'Ercole, Recchi,
\& Ciotti 1999), especially when a substantial SNIa rate (even though
not sufficient to extract the gas from the galaxy potential well) is
present. In these cases, transient cold filaments form and are
accreted on the center: in case of presence of a BH the situation
should evolve in a very complex way.

Another potential problem of the presented scenario is the value of
the temperature of the radiation emitted during the accretion events.
While the adopted value is higher than the mean gas temperature of the
ISM it is obvious that, in case of considerably lower radiation
temperature {\it Compton cooling} is expected, with a stronger
tendency towards a {\it radiation induced cooling flow}, as stated by
Nulsen \& Fabian (1999) (see Section 4.2.4).  As a consequence, one
should observe an even {\it higher} luminosity from the centers of
elliptical galaxies (although at low temperatures) than that predicted
by the simple energetic argument presented in the Introduction. A
possible solution to this problem could be a significant mass dropout
from the cooling flow, well outside the galaxy center. For values of
$\tx$ moderately smaller than assumed in the bulk of models presented
in this paper (but larger than $T_{\rm gas}$) one can, without further
computations, find the approximate solution by scaling $\tx$ down and
$\eps$ up by the same factor.

As a final comment on the present models, it seems to us unescapable
that in every case the presence of a massive BH at the center of
elliptical galaxies must produce a very strong feedback on the flow
itself, either of energetic nature, mainly via Compton heating (or
cooling, depending on the range of $\tx$, as discussed in this paper),
or via mechanical heating, as discussed in BT95.  It is important to
stress that both investigations, that given by BT95 and that presented
in CO97 and in this paper, are not in opposition to one another, but
complementary, demonstrating that in the presence of a massive BH the
possibility of a stationary cooling flow seems to be very remote; it
is instead more likely that a global equilibrium will be achieved
wherein the time--averaged heating and cooling rates are approximately
in balance.

Finally, we can speculate on the physical status of the gas in rich
galaxy clusters, assuming the presence of a massive BH at the center
of the cD galaxy. In this case a global degassing cannot be expected,
but the effect of the radiation feedback on the ICM should be still
strong, and the standard cooling flow scenario modified accordingly.
We point out that observations are available supporting the scenario
here presented, for example strong evidence of intermittent cooling
flows at the center of galaxy clusters (see, e.g., McNamara 1999,
Soker et al. 2000), and the failure of detection of the expected
amount of cold gas in the center of clusters (Miller, Bregman \&
Knezek 1999). And finally there is considerable evidence (see, e.g.,
Suginohara \& Ostriker 1998, Pen 1999, and references therein) for
additional source of entropy in the central regions of clusters.

\acknowledgments
We would like to thank James Binney, Piero Madau and Bohdan Paczynski
for useful discussions and advice, and Thierry Courvoisier for
providing us the spectral energy distribution of 3C 273. L.C. wishes
also to thank Giuseppe Bertin, Fabrizio Brighenti, Alberto Cappi,
Gabriele Ghisellini, Silvia Pellegrini, Alvio Renzini and Gianni
Zamorani for useful discussions. A particular acknowledgement is due
to Annibale D'Ercole for his advice on the original hydrodynamical
code.  The anonymous referees are thanked for useful comments that
improved the presentation of the paper. L.C. was supported by NSF
Grant AST9108103, by ASI, contract ASI-95-RS-152, and by MURST,
contract CoFin98. J.P. Ostriker was supported by NSF grant \#
94-24416.

\appendix

\section{The Input Physics}
We describe here the main ingredients entering the heating and cooling
terms $\lh$ and $\lrad$ in the hydrodynamical equations (14)-(16).
\subsection{Compton Heating and Cooling}

Under the assumption of spherical symmetry, the gas Compton heating 
(or cooling) for unit frequence at radius $r$ is given by:
\begin{equation}
\Delta\eco=-\nscat (\nu,r)\Delta\ega (\nu,T),
\end{equation}
where $\Delta\eco$ is the gas energy for unit volume gained (or lost)
by the gas at the frequence $\nu$ and at the radius $r$, and
$\Delta\ega$ is the energy variation of a photon of frequency $\nu$
interacting with an electron at the gas temperature $T(r)$.  The
number of photon-electron scatterings can be written as:
\begin{equation}
\nscat (\nu,r)=\ngamma (\nu,r)\times
{\sigknnu \nel (r)\over4\pi r^2}={\lbh (\nu,r)\Delta t\over h\nu}
\times{\sigknnu\nel (r)\over 4\pi r^2},
\end{equation}
where $\nel (r)$ is the electron number density, $\lbh (\nu,r)$ is the
BH luminosity at the radius $r$, and $\sigknnu=\sigt \tilde\sigknx$ is
the Klein--Nishina cross--section.
$\sigt\simeq 6.65\times 10^{-25}$ cm$^2$ is the Thomson cross section, 
$x=\nu h/\mel c^2$, and 
\begin{equation}
\tilde\sigknx = {3\over 4}\left\{ {1+x\over x^2}
\left [{2(1+x)\over 1+2x}-{\ln (1+2x)\over x}\right ]+{\ln (1 + 2x)\over 2x}-
{1+3x\over (1+2x)^2}\right\}
\end{equation}
(\cite{lang80}, p.68).  A simple approximation for
the energy transfer factor is\footnote{
This formula reproduces the well known relations $\Delta\ega\sim
1.5\kb T-\mel c^2 x+O(1/x)$ for relativistic photon energy ($h\nu\gg \mel
c^2$), and $\Delta\ega\sim 4\kb Tx-\mel c^2x^2+O(x^3)$ in the classical
limit.}:
\begin{equation}
\Delta\ega(\nu,T)={4\kb Tx(1+3x^2/8)\over 1 + x^3}-
                  {\mel c^2 x^2(1+x^2)\over 1+x^3}.
\end{equation}
After substitution of equations (A2) and (A4) in equation (A1) one obtains 
\begin{equation}
\left ({\pd {E} t}\right)_{\rm C}=-E {8\over 3}{\nel (r)\over\nt (r)}
{\lbh(\nu,r)\over \mel c^2}{\sigknx\over 4\pi r^2}\Delta (x,T),
\end{equation}
where $\Delta (x,T)\equiv\Delta E_{\gamma}(\nu,r)/4\kb T x$.
Note that $\lbh(\nu,r)$ satisfy the continuity equation 
\begin{equation}
{\partial \lbh(\nu,r)\over\partial r}=-4\pi r^2\left ({\pd E t}
\right )_{\rm C}.
\end{equation}
We speed up the numerical simulations assuming a {\it gray}
absorption, i.e., at any radius $\lbh(\nu,r)=\fznu\lbh(r)$.
With this choice, and after integration of equation (A5) over all 
frequencies, one has:
\begin{equation}
\left ({\pd E t}\right )_{\rm C}=-E{8\over 3}{\nel (r)\over\nt (r)}
{\sigt\over 4\pi r^2}{\lbh (r)\over \mel c^2}
\cx\left [1-{\tx\over T(r)}\right ],
\end{equation}
where
\begin{equation}
\cx=\int _0^{\infty}
{(1+3x^2/8)\tilde\fz (x)\tilde\sigknx\over 1 + x^3}\;dx =0.137,
\end{equation}
and the spectral temperature is given by:
\begin{equation}
\tx={\mel c^2\over 4\kb\cx}\int _0^{\infty}{x(1+x^2)\tilde\fz (x)
\tilde\sigknx\;dx \over 1+x^3}=1.04\times 10^9\quad {\rm K}.
\end{equation}
The two numerical values are obtained using $\fz$ as given by equation
(10).  The radial dependence of $\lbh$ is now obtained by integrating
equation (A6) over all the frequencies, by using equation (A7), and
finally integrating the resulting differential equation:
\begin{equation}
\lbh(r)=\lbh (0)\exp\left[{8\over 3}{\sigt\cx\over \mel c^2}
\int_0^rE{\nel\over\nt}\left (1-{\tx\over T}\right )\;dr\right ].
\end{equation}
$\lbh(0)$ is the bolometric luminosity emitted by the accreting
material on the central BH, as given by equation (1). Equation (A7)
can now be integrated with respect to time.

\subsection{Photoionization Heating}

We consider three photoionization processes, one for the hydrogen and
two for the helium. For each of the processes an 
equation similar to equation (A1) holds:
\begin{equation}
\Delta\eph=-\nph(\nu>\nui,r)\Delta\ega(\nu,\nui),
\end{equation}
where $\nui$ is the ionization frequency of HI, HeI and HeII. Moreover,
\begin{equation}
\nph (\nu>\nui,r)=\ngamma (\nu >\nui,r)\times {\sigi (\nu)\nei (r)
\over 4\pi r^2}={\lbh (\nu >\nui,r)\Delta t\over h\nu}\times
{\sigi (\nu)\nei (r)\over 4\pi r^2},
\end{equation}
where the ionization energies are
$(h\nuh,h\nuhe,h\nuhee)= (13.598,24.587,54.416)$ eV (\cite{lang80},
p.246), $\sigi (\nu)=\sigma_{\rm i\circ}\tilde\sigi (\nu/\nui)$ are
the photoionization cross--sections, and $\nei(r)$ is the number (per
unit volume) of HI, HeI, and HeII atoms, respectively.  For HI and HeII
atoms in the non--relativistic regime,
\begin{equation}
\tilde\sigi (y)={1\over y^4}
{\exp[-4{\rm arctg}(\sqrt{y-1})/\sqrt{y-1}]\over 1-\exp(-2\pi/\sqrt{y-1})},
\end{equation}
and for HeI
\begin{equation}
\tilde\sigi (y)=1.66 y^{-2.05}-0.66 y^{-3.05},
\end{equation}
with $(\sighc,\sighec,\sigheec)=(3.441\times 10^{-16},7.83\times
10^{-18}, 8.5992\times 10^{-17})$ cm$^{-2}$, (\cite{oster74}, p.15;
\cite{lang80}, p.469).  For the exchanged energies we use the simple
approximation $\Delta\ega=h\nui-h\nu$, and by integrating equation 
(A11) over all the frequencies three constants are derived:
\begin{equation}
C_{\rm i}={\nui\over\nut}\int_1^{\infty}\tilde\fz \left({x\nui\over\nut}\right)
\left(1-{1\over x}\right)\tilde\sigi (x)\;dx,
\end{equation}
with $(\ch,\che,\chee)=(1.55\times 10^{-6},2.99\times 10^{-4}, 3.09\times
10^{-6})$.
Summing the rates for the three processes we finally obtain:
\begin{equation}
\left ({\pd E t}\right )_{\rm ph}={\lbh(r)\over 4 \pi r^2}
\sum_{i=1}^3 C_{\rm i}\sigma_{\rm i\circ}n_{\rm i}(r).
\end{equation}

\subsection{Bremsstrahlung Recycling}

The luminosity emitted per unit volume and per unit frequence over the
solid angle by \brem radiation from a gas at temperature $T(r)$ and atomic
number $Z_{\rm i}$, is given by 
\begin{equation}
\dot\eb (\nu,r)={2^{11/2}(\pi /3)^{3/2}Z_{\rm i}^2 g_{\rm ff}\qe^6\over\mel^2
c^4}\nel\nei\left ({\mel c^2\over \kb T}\right )^{1/2}\exp{(-h\nu/\kb T)}.
\end{equation}
The bolometric \brem luminosity is then given by:
\begin{equation}
\dot\eb (r)={2^{11/2}(\pi /3)^{3/2}Z_{\rm i}^2\overline g \qe^6
\over h \mel c^2}\nel\nei\left ({\kb T\over\mel c^2}\right )^{1/2},
\end{equation}
(\cite{lang80}, pp.46-48). 
Thus, the (normalized) \brem spectral distribution is:
\begin{equation}
\dot\eb (\nu,r)=\fb(\nu)\dot\eb (r), \quad
\fb (\nu)={g_{\rm ff}\over \overline g}{h\over \kb T}\exp{(-h\nu/\kb T)}.
\end{equation}
The equation for the energy transfer from photons to electrons 
is obviously analogous to equation (A1), where now:
\begin{equation}
\nscat (r,\nu)={\lbr (r,\nu)\Delta t\over h\nu}\times
{\nel (r)\sigknnu\over 4\pi r^2}.
\end{equation}
The total luminosity emitted by \brem inside the region of radius $r$ 
is given by:
\begin{equation}
\lbr (r,\nu)=4\pi\int_0^r r'^2\dot\eb (r',\nu)\;dr'.
\end{equation} 
In order to speed up the computation we use also for the \brem emission
the approximation of gray atmosphere and $g_{\rm ff}/\overline g=1$.
In this case, after integration over the frequency, we obtain:
\begin{equation}
\left ({\pd E t}\right )_{\rm Br}=-E{8\over 3}{\nel\over\nt}
                                   {\sigt\over 4\pi r^2}
4\pi\int_0^r r'^2\dot\eb (r')F(r',r,y)\;dr',
\end{equation}
where, using eqs. (A4)-(A19), and defining $t(r')\equiv \kb T(r')/\mel c^2$:
\begin{equation}
F(r',r)=\int_0^{\infty}\tilde\sigkn [t(r') y]\exp{(-y)}\left[
          {1+3t(r')^2y^2/8\over 1+t(r')^3y^3}-
          {T(r')y\over 4T(r)}{1+t(r')^2y^2\over 1+t(r')^3y^3}
          \right]\;dy .
\end{equation}

\subsection{Effective Cooling Function}

The radiative losses (i.e., the cooling rate per unit volume) 
are described by the function $\lrad=\nel n_{\rm p}
\Lambda(T,\Xi )$, where 
\begin{equation}
\Lambda (T,\Xi)=\cases {[\Lambda _{\circ}(T) - B_{\circ}(T)]
\exp (-100\;\Xi)\;+B_{\circ}(T)
& for $T\leq 10^8$ K\cr
B_{\circ}(T) &for $T\geq 10^8$ K.\cr}
\end{equation}
The function 
$\Lambda _{\circ}(T)$ (erg cm$^3$ s$^{-1}$) is given for example 
by Mathews \& Bregman (1978) (after correction of a wrong sign in their
formula):
\begin{equation}
\Lambda _{\circ}(T)=\cases{5.3547\times 10^{-27}T& for $10^4  K\leq T\leq
1.3\times 10^5$ K,\cr
2.1786\times 10^{-18}T^{-0.6826}+2.7060\times 10^{-47}T^{2.976}&
for $1.3\times 10^5\leq T\leq 10^8$ K;\cr}
\end{equation}
when $T\geq 10^8$ K (the \brem branch)
$B_{\circ}(T)=\Lambda_{\circ}(10^8)
\sqrt {T/10^8}$.
The ionization parameter $\Xi$ is introduced in order to describe 
the effects on the radiative losses by the hydrogen photoionization, 
and  
\begin{equation}
\Xi (r)= {n_{\gamma}(\nu\geq\nuh)\over n_{\rm H}},
\end{equation}
where $n_{\gamma}$ is the number density of photons with an energy
greater than the hydrogen ionization energy. The parameter $\Xi$ can be
easily derived in case of gray atmosphere as
\begin{equation}
n_{\gamma}(\nu \geq\nuh)=\int_{\nuh}^{\infty}
{\lbh\fznu\over 4\pi r^2 c h \nu}d\nu=
{\lbh (r)\over 4 \pi r^2 \mel c^3}\int_{\nuh/\nut}^\infty 
\tilde\fz (x){dx,\over x},
\end{equation}
where the dimensionless integral equals 47.813 when using the spectral
energy distribution given in equation (10).

\subsection{The Effective Gravitational Field}

Photons emitted by the BH interacts with the surrounding gas not only
by exchanging energy but also exchanging momentum. This results in a
modification of the effective gravity in the momentum equation (15):
\begin{equation}
\gef (r)=-{GM(r)\over r^2}\left [
1-{\nel (r)\int _0^{\infty}\sigknnu \lbh (\nu,r)\;d\nu\over
4\pi G\rho (r)M(r)c}\right ].
\end{equation}
Note that, by imposing the vanishing of the effective gravitational field
and assuming $M(r)=\mbh$, one obtains the classical expression for the
Eddington luminosity:
\begin{equation}
\ledd={4\pi G c\mbh\over\sigt C_{\rm Edd}}{\rho (r)\over\nel (r)},
\end{equation}
where for the frequency distribution (10) one obtains
$C_{\rm Edd}=0.214$.

\section{Numerical evaluation of $\lbh$ with $\tac >0$}

We describe here the numerical technique used to compute numerically
$\lbh$.

Let $R_1$ the nearest grid--point to the center, $\rho_1$ and $v_1$
the density and the velocity of gas at this point, and
$t_{i+1}$ the present value of the time.
We use the following approximation for the emitted luminosity:
\begin{equation}
\lbh (t_{i+1})=
{\eps c^2\over\tac}\int_0^{t_{i+1}}\fle (t')e^{-(t_{i+1}-t')/\tac}\;dt',
\end{equation}
where $\fle (t')=-4\pi R^2_1\rho_1 v_1$ if $v_1<0$ and 0 otherwise.
In this way we qualitatively describe the smooth release of gas 
from the accretion disk to the BH. For example, 
for a stationary accretion with $\fle=\fle(0)$ one obtains 
$\lbh(t)=\eps c^2 \fle (0)[1-\exp (-t/\tac)]$, and for an impulsive accretion
$\fle=M_{\rm accr} \delta (t)$, $\lbh(t)=
\eps c^2 (M_{\rm accr}/\tac)\exp (-t/\tac)$.
It is easy to prove the following identity:
\begin{equation}
\lbh(t_{i+1})=\lbh (t_i)e^{-(t_{i+1}-t_i)/\tac}
+{\eps c^2 e^{-t_{i+1}/\tac}\over\tac}\int_{t_i}^{t_{i+1}}\fle (t')
e^{-(t_{i+1}-t')/\tac}\;dt'.
\end{equation}
In the integration of the hydrodynamical 
equations, the integral over the time-step of equation
(B2) is required. Changing
order of integration on $\Delta E_{\rm BH}=\int_{t_{\rm i}}^{t_{\rm i+1}}\lbh
(t)\,dt$ one obtains:
\begin{equation}
\Delta E_{\rm BH}=\tac[\lbh (t_i)-\lbh (t_{i+1})]+
\eps c^2\int_{t_i}^{t_{i+1}}\fle (t')e^{-(t_{i+1}-t')/\tac}\;dt'.
\end{equation}
During each time--step the function $\fle$ is defined as the linear
interpolation between the initial and final time: the integral in
equation (B3) can be explicitly calculated, and the computer time
required for its evaluation is negligible.

\clearpage
\begin{deluxetable}{crrrrrrrr}
\footnotesize
\tablecaption{Overall properties of the presented models. \label{tbl-1}}
\tablewidth{0pt}
\tablehead{
           \colhead{Model \#} &
           \colhead{$\eps$}  &
           \colhead{$\vtsn$} &
           \colhead{$\mr$} &
           \colhead{$\tcc$} &
           \colhead{ADAF} &
           \colhead{$\Delta\mbh$} &
           \colhead{$\fBH$}
          }
\startdata
0                  &         0              &         2/3         &      
7.8                &         9.05           &         No          &
$1.5\times 10^{10}$&         $-$            \nl\nl

1                  &         0.1            &         2/3         &      
7.8                &         9.05           &         No          &
$1.0\times 10^8$   & $1.0\times 10^{-4}$    \nl
2                  &         0.01           &         2/3         &      
7.8                &         9.05           &         No          &
$3.1\times 10^8$   & $2.7\times 10^{-4}$    \nl
3                  &         0.001          &         2/3         &
7.8                &         9.05           &         No          &
$4.3\times 10^8$   & $3.8\times 10^{-4}$    \nl\nl

4                  &         0.1            &         2/3         &      
10                 &         4.24           &         No          &
$5.8\times 10^8$   & $2.1\times 10^{-4}$    \nl
5                  &         0.01           &         2/3         &      
10                 &         4.24           &         No          &
$7.0\times 10^8$   & $2.6\times 10^{-4}$    \nl
6                  &         0.001          &         2/3         &      
10                 &         4.24           &         No          &
$6.9\times10^8$    & $2.5\times 10^{-4}$    \nl\nl

7                  &         0.1            &         1/4         &      
7.8                &         0.25           &         No          &
$3.5\times 10^9$   & $1.5\times 10^{-3}$    \nl
8                  &         0.01           &         1/4         &      
7.8                &         0.25           &         No          &
$4.9\times 10^9$   & $1.8\times 10^{-3}$    \nl
9                  &         0.001          &         1/4         &      
7.8                &         0.25           &         No          &
$5.0\times 10^9$   & $1.7\times 10^{-3}$    \nl\nl

10                 &         0.1            &         1/4         &      
3.0                &         0.25           &         No          &
$2.7\times 10^9$   & $1.4\times 10^{-3}$    \nl
11                 &         0.01           &         1/4         &      
3.0                &         0.25           &         No          &
$4.4\times 10^9$   & $1.9\times 10^{-3}$    \nl
12                 &         0.001          &         1/4         &      
3.0                &         0.25           &         No          &
$4.6\times 10^9$   & $2\times 10^{-3}$      \nl\nl

13                 &         0.1            &         2/3         &      
7.8                &         9.05           &         Yes         &
$2.1\times 10^8$   & $1.2\times 10^{-4}$    \nl
14                 &         0.01           &         2/3         &      
7.8                &         9.05           &         Yes         &
$4.4\times 10^9$   & $1.3\times 10^{-3}$    \nl
15                 &         0.001          &         2/3         &      
7.8                &         9.05           &         Yes         &
$1.5\times 10^{10}$& $3.3\times 10^{-3}$    \nl\nl

16                 &         0.1            &         2/3         &      
7.8                &         9.05           &         No          &
$1.6\times 10^9$   & $5.3\times 10^{-2}$    \nl
17                 &         0.1            &         2/3         &      
7.8                &         9.05           &         No          &
$1.6\times 10^9$   & $5.4\times 10^{-2}$    \nl\nl

1r                 &         0.1            &         2/3         &      
7.8                &         9.05           &         No          &
$4.5\times 10^8$   & $4.1\times 10^{-4}$    \nl
2r                 &         0.01           &         2/3         &      
7.8                &         9.05           &         No          &
$4.0\times 10^8$   & $3.5\times 10^{-4}$    \nl
1rr                &         0.1            &         2/3         &      
7.8                &         9.05           &         No          &
$5.9\times 10^8$   & $5.3\times 10^{-4}$    \nl\nl

\enddata
\tablecomments{All models in the table have $\lb =5\times 10^{10}\lsol$,
               $\ssc =280$ km s$^{-1}$, $\rcs =350$ pc, $\rt =63$ kpc.
               The amount of dark matter is a free parameter, but its
               scale--length is fixed to $\rch=4.2\rcs$. $\Delta\mbh$
               is the accreted mass by the central BH in solar units
               at the end of the simulation, when $t =15$ Gyr. Models
               \#16,17 simulate the presence of a geometrically thin,
               optically thick accretion disk.}
\end{deluxetable}
\clearpage
\begin{deluxetable}{crrrrr}
\footnotesize
\tablecaption{Time averaged X--ray luminosity. \label{tbl-2}}
\tablewidth{0pt}
\tablehead{
           \colhead{Model \#} &
           \colhead{$\eps$}  &
           \colhead{$\tx$} &
           \colhead{ADAF} &
           \colhead{$<\lx >$}
          }
\startdata
0                  &         0              &        $-$          &      
       No          &        100             \nl\nl

1                  &         0.1            &        $10^9$       &      
       No          &        7.6             \nl
2                  &         0.01           &        $10^9$       &      
       No          &        13              \nl
3                  &         0.001          &        $10^9$       &      
       No          &        15              \nl\nl

13                 &         0.1            &        $10^9$       &      
       Yes         &         5.7            \nl
14                 &         0.01           &        $10^9$       &      
       Yes         &         25             \nl
15                 &         0.001          &        $10^9$       &      
       Yes         &         30             \nl\nl

16                 &         0.1            &        $10^9$       &      
       No          &         30             \nl
17                 &         0.1            &        $10^9$       &      
       No          &         30             \nl\nl

1r                 &         0.1            &        $10^8$       &      
       No          &         10.6           \nl
2r                 &         0.01           &        $10^8$       &      
       No          &         13.5           \nl
1rr                &         0.1            &     $5\times 10^7$  &      
       No          &         15.3           \nl\nl

\enddata
\tablecomments{Time averaged X--ray luminosity $<\lx>$ (for $t>\tcc$) 
               in the energy range $0.5 - 4.5$ keV, in units of
               $10^{40}$ erg s$^{-1}$, for all models in Table 1 with
               the same structural parameteres.}
\end{deluxetable}
\clearpage
\begin{deluxetable}{crrrrrrrr}
\footnotesize
\tablecaption{Overall properties of models presented in Section 4.4. 
\label{tbl-3}}
\tablewidth{0pt}
\tablehead{
           \colhead{Model \#} &
           \colhead{$\lb$}  &
           \colhead{$\ssc$} &
           \colhead{$\tcc$} &
           \colhead{$\tx$} &
           \colhead{$\fBH$} &
           \colhead{$\Delta\mbh$} &
           \colhead{$< \lx >$}
          }
\startdata
18                 &         3              &         200         &      
$-$                &         $10^9$         &         $-$         &
0                  &         1.4            \nl

19                 &         3              &         210         &      
9.5                &         $10^9$         &$4.7\times 10^{-4}$  &
$6.5\times 10^7$   &         0.07           \nl

20                 &         3              &         220         &      
4.6                &         $10^9$         &$5.6\times 10^{-4}$  &
$2.9\times 10^8$   &         0.24           \nl

21                 &         4              &         210         &
9.6                &         $10^9$         &$1.7\times 10^{-4}$  &
$4.8\times 10^7$   &         0.08           \nl

22                 &         4              &         220         &      
7.2                &         $10^9$         &$2.7\times 10^{-4}$  &
$1.2\times 10^8$   &         0.27           \nl

21r                &         4              &         210         &      
9.6                &    $2\times 10^8$      &$6.2\times 10^{-4}$  &
$1.4\times 10^8$   &         0.2            \nl

\enddata
\tablecomments{All models in the table have $\eps=0.1$, $\mr =2$, and
               $\vtsn =1/4$.  Galaxy optical luminosities are in units
               of $10^{10}\lsol$, and central velocity dispersions in
               km s$^{-1}$.  Time averaged X--ray luminosity $<\lx>$
               (for $t>\tcc$, except for wind model \#18), is in units
               of $10^{40}$ erg s$^{-1}$.}
\end{deluxetable}
\clearpage

\clearpage

%
%

\clearpage

\figcaption[]{The time evolution of mass and energy budget of model \#1. 
              Masses are given in solar masses, mass rates in solar
              masses per year, and luminosity in erg per second.  In
              panel a, the mass of gas inside $\rt$ ($\mgas$, dotted
              line), and the mass accreted by the BH ($\msink$, solid
              line) are shown.  In panel b the mass return rate from
              the stellar population ($\dot\mast$) is represented by
              the dotted line, the rate of mass loss at the truncation
              radius $\rt$ ($\dot\mout$) by the dashed line, and
              finally the BH accretion rate ($\mdot$) by the solid
              line.  In panel c the short dashed and long dashed lines
              represent $\lsn$ and $\lgm$, respectively. The solid
              line represents $\lx$, the observable X--ray gas
              emission calculated inside $\rt$ and in the range
              0.5--4.5 keV.  $\lbh$ is represented by the dotted
              line. In panel d we show the time evolution of the gas
              temperature (solid line, scale on the left axis) and
              density (dotted line, scale on the right axis) at 20 pc
              from the center.  \label{fig1}}

\figcaption[]{Time evolution of $\lx$ for model \#1 (solid line) and for  
              model \#0 identical to model \#1 (dotted line), but
              without the feedback from the central BH.  The
              time--averaged luminosity of the two models \#0,1 are
              respectively $10^{42}$ and $7.6\times 10^{40}$ erg
              s$^{-1}$.  \label{fig2}}

\figcaption[]{Evolution of several quantities for model \#1 
              in the time interval 9.04 -- 9.8 Gyr.  The time sampling
              of the plotted quantities is $10^4$ yr.  The linestyle
              and physical units are the same as in Fig.1.  Note the
              rich temporal sub--structure of each accretion event;
              moreover, the opposite trend of the temperature (solid
              line) and density (dotted line) in the central regions
              of the flow is apparent in panel c. Peak luminosities,
              not shown, reach $10^{47}$ erg s$^{-1}$. \label{fig3}}

\figcaption[]{Evolution of the temperature, density, 
              and velocity profiles (upper, middle, and lower panels,
              respectively), of model \#1.  In each panel the time
              sequence is: solid, dotted, short-dashed, long-dashed,
              dotted-dashed, and the (constant) time interval is 10
              Myr. The solid lines in the three columns are the
              profiles at 9.01, 9.06, and 9.2 Gyrs, i.e., just before
              (column 1) and just after (column 2) $\tcc$, while the
              time sequence in column 3 starts at 9.2 Gyrs, after a
              strong central burst (cfr. with Fig.3c).  Note that the
              ordinate scales in the first and third temperature and
              density panels cover identical ranges, allowinwg a
              direct comparison of the various profiles.
              \label{fig4}}

\figcaption[]{The evolution of $\lbh$ and $\lx$ for models \#1,2,3,
              from 9 Gyr ($\simeq\tcc$) to 15 Gyr, sampled at time
              intervals of 1 Myr. The accretion efficiency in the
              three models is $\eps=(0.1,0.01,0.001)$, respectively.
              The various linetypes are the same as in
              Fig.1c. \label{fig5}}

\figcaption[]{The evolution of $\lbh$ and $\lx$ for models \#4,5,6,
              from 4.2 ($\simeq\tcc$) to 15 Gyr, sampled at time
              intervals of 1 Myr. The accretion efficiency in the
              three models is again $\eps=(0.1,0.01,0.001)$,
              respectively, but the dark matter halo is more massive
              (see Table 1).  The various linetypes are the same as in
              Fig.1c. \label{fig6}}

\figcaption[]{The time evolution of $\lbh$ and $\lx$ for models \#7,10,17,
              from 9 Gyr to 9.5 Gyr (upper, middle, and lower panels,
              respectively).  All three model are characterized by
              $\eps=0.1$, and the various linetypes corresponds to
              those described in Fig.1. In model \#17, $\tac=10^4\tx$.
              \label{fig7}}

\figcaption[]{The evolution of $\lbh$ and $\lx$ for the ``ADAF'' models 
              \#13,14,15, from 9 Gyr ($\simeq\tcc$) to 15 Gyr, sampled at
              time intervals of 1 Myr. The accretion efficiency in the
              three models is $\eps=(0.1,0.01,0.001)$, respectively.
              The various linetypes are the same as in
              Fig.1c. \label{fig8}}

\figcaption[]{The evolution of $\lbh$ (dotted line) and $\lx$ (solid line) 
              of models \#21 (upper panel), and \#21r (lower panel), 
              sampled at time intervals of 1 Myr. 
              \label{fig9}}

\figcaption[]{The X--ray surface brightness profiles (in arbitrary units) 
              of models \#1 (upper panel) and model \#7 (lower panel)
              at ten different times, sampled at time intervals of 0.5
              Gyr, and starting from 9.5 Gyr. The heavy solid lines
              are the X--ray surface brightness profiles of the
              corresponding models without central BH, at arbitrary time 
              $t=12$ Gyr.
              \label{fig10}}

\clearpage


\begin{thebibliography}{}

\bibitem[Arimoto et al.\ 1997]{ari97} Arimoto, N., Matsushita, K., 
        Ishimaru, Y., Ohashi, T., \& Renzini, A. 1997,
        \apj, 477, 128
\bibitem[Awaki et al.\ 1991]{aetal91} Awaki, H., Koyama, K., Kunieda, H., 
        Takano, S., Tawara, Y., \& Ohashi, T. 1991, 
        \apj, 366, 88
\bibitem[Awaki et al.\ 1994]{aetal94} Awaki, H. et al. 1994, \pasj, 46, L65
\bibitem[Bahcall et al.\ 1999]{bahcetal99} Bahcall, N.A., Ostriker, J.P., 
        Perlmutter, S., \& Steinhardt, P.J. 1999,
        {\it Science}, 284, 1481
\bibitem[Bertin \& Toniazzo 1995]{bto95} Bertin, G., \& Toniazzo, T. 1995,
        \apj, 451, 111
\bibitem[Binney \& Tabor 1995]{bt95} Binney, J., \& Tabor, G. 1995,
        \mnras, 276, 663 (BT95)
\bibitem[Brighenti \& Mathews 1996]{bm96} Brighenti, F., \& Mathews, W.G.
        1996, \apj, 470, 747
\bibitem[Brighenti \& Mathews 1999]{bm99} Brighenti, F., \& Mathews, W.G.
        1999, preprint, (astro-ph/9910556)
\bibitem[Buote 1999]{bu99} Buote, D.A. 1999, \mnras, 309, 685
        \mnras, 296, 977
\bibitem[Buote \& Fabian 1998]{bf98} Buote, D.A., \& Fabian, A.C. 1998,
        \mnras, 296, 977
\bibitem[Cappellaro et al.\ 1997]{cap97} Cappellaro, E., Turatto, M.,
        Tsvetkov, D.Y., Bartunov, O.S., Pollas, C., Evans, R., \& Hamuy, M.
        1997, A\&A, 322, 431
\bibitem[Canizares et al.\ 1987]{cft87} Canizares, C.R., Fabbiano, G., 
        \& Trinchieri, G. 1987, \apj, 312, 503
\bibitem[Ciotti 1993]{c93} Ciotti, L. 1993,
        Ph.D. Thesis, Astronomy Dept., University of Bologna, Italy
\bibitem[Ciotti et al. 1991]{cdpr91} Ciotti, L., D'Ercole, A., Pellegrini, S.,
        \& Renzini, A., \apj, 376, 380 (CDPR)
\bibitem[Ciotti \& Ostriker 1997]{co97} Ciotti, L., \& Ostriker, J.P. 1997,
        \apjl, 487, L105 (CO97)
\bibitem[Cowie et al.\ 1978]{cos78} Cowie, L.L., Ostriker, J.P., \&
        Stark, A.A. 1978, \apj, 226, 1041
\bibitem[D'Ercole \& Ciotti 1998]{dc98} D'Ercole, A., \& Ciotti, L. 1998,
        \apj, 494, 535
\bibitem[D'Ercole et al.\ 1999]{drc99} D'Ercole, A., Recchi, S., \& Ciotti, L.
        1999, preprint, (astro-ph/9910142)
\bibitem[Eskridge et al.\ 1995]{esk95} Eskridge, P.B., Fabbiano, G., \& Kim,
        D.W. 1995, \apjs, 97, 141
\bibitem[Fabbiano 1989]{fab89}Fabbiano, G. 1989, \araa, 27, 87
\bibitem[Fabian et al.\ 1984]{fnc84} Fabian, A.C., Nulsen, P.E.J., \& 
        Canizares, C.R. 1984, Nature, 311, 733
\bibitem[Fabian \& Rees 1995]{fr95} Fabian, A.C., \&  Rees, M.J. 1995,
        \mnras, 277, L55
\bibitem[Ferrarese \& Merritt 2000]{fm2000} Ferrarese, L., \& Merritt, D. 2000,
        \apj, 539, L9
\bibitem[Forman et al.\ 1979]{form79} Forman, W., Schwarz, J., Jones, C., 
        Liller, W., \& Fabian, A. 1979, \apjl, 234, L27
\bibitem[Forman et al.\ 1985]{fjt85} Forman, W., Jones, C., \& Tucker, W. 
        1985, \apj, 293, 102
\bibitem[Fukugita \& Peebles 1999]{fp99} Fukugita, M., \& Peebles, P.J.E. 
        1999, preprint (astro-ph/9906036)
\bibitem[Gebhardt et al.\ 2000]{geb2000} Gebhardt et al. 2000, \apj, 
        539, 13
\bibitem[Giacconi et al.\ 2000]{giac2000} Giacconi et al. 2000, preprint,
        (astro-ph/0007240)
\bibitem[Harms et al.\ 1994]{har94} Harms, R.J., et al.. 1994, \apjl, 435,
        L35
\bibitem[Ikebe et al.\ 1992]{ike92} Ikebe, Y., et al. 1992, 
        \apjl, 384, L5 
\bibitem[Kim et al. 1996]{kim96} Kim, D.W., Fabbiano, G., Matsumoto, H., 
        Koyama, K., \& Trinchieri, G. 1996,
        \apj, 468, 175
\bibitem[King 1972]{k72} King, I. 1972, \apjl, 174, L123
\bibitem[Kormendy \& Ho 2000]{kh00} Kormendy, J., \& Ho, L.C. 2000, 
        preprint, (astro-ph/0003268)
\bibitem[Lang 1980]{lang80} Lang, K.R. 1980,
        Astrophysical Formulae, Springer-Verlag, New York 
\bibitem[Loewenstein et al.\ 1994]{lmt94} Loewenstein, M., Mushotzky, R.F.,
        Tamura, T., Ikebe, Y., Makishima, K., Matsushita, K., Awaki, H., \&
        Serlemitsos, P.J. 1994, \apjl, 436, L75
\bibitem[Lu \& Yu 1999]{ly99} Lu, Y., \& Yu, Q. 1999,
        preprint, (astro-ph/9911289)
\bibitem[Macomb et al.\ 1995]{mac95} Macomb, D.J., et al. 1995, \apjl,
        449, L99
\bibitem[Madau \& Efstathiou 1999]{me1999} Madau, P., \& Efstathiou, G. 1999,
        \apjl, 517, L9
\bibitem[McNamara 1999]{mcn99} McNamara, B.R. 1999, 
        preprint, (astro-ph/9911129)
\bibitem[McNaron-Brown et al.\ 1997]{mcna97} McNaron-Brown, K., Johnson, C.D., 
        Dermer, C.D., \& Kurfess, J.D. 1997, \apjl, 474, L85
\bibitem[Magorrian et al.\ 1998]{mag98} Magorrian, J., et al. 1998, \aj,
        115, 2285
\bibitem[Maraschi et al.\ 1994]{mara94} Maraschi, L., et al. 1994, \apjl, 
        435, L91
\bibitem[Mathews \& Bregman 1978]{mb78} Mathews, W.G., \& Bregman, J.N. 
        1978, \apj, 224, 308
\bibitem[Mathews \& Brighenti 1998]{mb98} Mathews, W.G., \& Brighenti, F.
        1998, \apjl, 493, L9
\bibitem[Mathews \& Brighenti 1999]{mb99} Mathews, W.G., \& Brighenti, F.
        1999, \apj, 526, 114
\bibitem[Matsumoto et al.\ 1997]{mat97} Matsumoto, H., Koyama, K., Awaki, H., 
        Tsuru, T., Loewenstein, M., \& Matsushita, K. 1997, 
        \apj, 482, 133
\bibitem[Miller et al.\ 1999]{mill99} Miller, E.D., Bregman, J.L., \&
        Knezek, P.M. 1999,
        preprint, (astro-ph/9911021)
\bibitem[Mushotzky 1999]{mu99} Mushotzky, R.F. 1999, in The Hy Redshift
        Universe, ASP Conference Series, vol. 193, p.323, A.J. Bunker 
        \& W.J.M. van Breugel, eds.
\bibitem[Mushotzky et al.\ 1994]{mla94} Mushotzky, R.F., Loewenstein, M., 
        Awaki, H., Makishima, K., Matsushita, K., \& Matsumoto, H. 1994,
        \apjl, 436, L79
\bibitem[Mushotzky et al.\ 2000]{metal00} Mushotzky, R.F., Cowie, L.L., 
        Barger, A.J., \& Arnaud, K.A. 2000, 
        Nature, 404, 459
\bibitem[Narayan \& Yi 1995]{ny95} Narayan, R., \& Yi, I. 1995,
        \apj, 452, 710 (NY95)
\bibitem[Nulsen et al.\ 1984]{nsf84} Nulsen, P.E.J., Stewart, G.C., \&
        Fabian, A.C. 1984, \mnras, 208, 185
\bibitem[Nulsen \& Fabian.\ 1999]{nf99} Nulsen, P.E.J., \& Fabian, A.C.       
        1999, preprint, (astro-ph/9908282)
\bibitem[Ohashi et al.\ 1990]{oha90} Ohashi, T., et al. 1990,
        in Windows on Galaxies, eds. G. Fabbiano, J.A. Gallagher, \& 
        A. Renzini (Dordrecht: Kluwer), 243
\bibitem[Osterbrock 1974]{oster74} Osterbrock, D.E. 1974, Astrophysics
        of Gaseous Nebulae, W.H. Freeman and Company, San Francisco
\bibitem[Ostriker et al.\ 1976]{omccwy76} Ostriker, J.P., McCray, R.,
        Weaver, R., \& Yahil, A. 1976, \apjl, 208, L61 
\bibitem[Park \& Ostriker 1998]{po98} Park, M.-G., \& Ostriker, J.P. 1998,
        Adv. Space Res., 22, p.951
\bibitem[Park \& Ostriker 1999]{po99} Park, M.-G., \& Ostriker, J.P. 1999,
        preprint, (astro-ph/9901243)
\bibitem[Pellegrini \& Ciotti 1998]{pc98} Pellegrini, S., \& Ciotti, L. 1998,
        A\&A, 333, 433
\bibitem[Pellegrini 1999]{p99} Pellegrini, S. 1999,
        preprint, (astro-ph/9909458)
\bibitem[Pen 1999]{pen99} Pen, U. 1999,
        \apjl, 510, L1
\bibitem[Quataert \& Narayan 1999]{qn99} Quataert, E., \& Narayan, R. 1999,
        preprint, (astro-ph/9908199)
\bibitem[Raymond \& Smith 1977]{rs77} Raymond, J., \& Smith, B. 1977,
        ApJS 35, 419
\bibitem[Rees 1984]{r84} Rees, M.J. 1984, \araa,
        22, 471
\bibitem[Renzini et al.\ 1993]{rcdp93} Renzini, A., Ciotti, L., D'Ercole, A.,
        \& Pellegrini, S. 1993, \apj, 419, 52
\bibitem[Richstone 1998]{rich98} Richstone, D.O. 1998, in IAU Symp. 184,
        The Central Region of the Galaxy and Galaxies, ed. Y. Sofue, in press
\bibitem[Sarazin \& Ashe 1989]{sa89} Sarazin, C.L. \& Ashe, G.A. 
        1989, \apj, 345, 22 
\bibitem[Sarazin \& White 1987]{sw87} Sarazin, C.L. \& White, R.E.III 
        1987, \apj, 320, 32
\bibitem[Sarazin \& White 1988]{sw88} Sarazin, C.L. \& White, R.E.III 
        1988, \apj, 331, 102
\bibitem[Serlemitsos et al.\ 1993]{ser93} Serlemitsos, P.J., Loewenstein, M.,
        Mushotzsky, R., Marshall, F., \& Petre, R. 1993,
        \apj, 413, 518
\bibitem[Shakura \& Sunyaev 1973]{ss73} Shakura, N.I., \& Sunyaev, R.A.
        1973, A\&A, 24, 337
\bibitem[Shapiro 1973]{shap73} Shapiro, S.L.
        1973, \apj, 180, 531
\bibitem[Soker et al.\ 2000]{sok2000} Soker, M.N., White, R.E.III, David, L.P.,
        McNamara, B.R.
        2000, preprint (astro-ph/0009173)
\bibitem[Suginohara \& Ostriker 1998]{so98} Suginohara, T., \& Ostriker, J.P.
        1998, \apj, 507, 16
\bibitem[Tammann 1982]{tam82} Tammann, G. 1982, Supernovae: A Survey of 
        Current Research, ed. M. Rees and R. Stoneham, Dordrecht: Reidel, 371
\bibitem[T\"urler et al. 1999]{tur99} T\"urler, M., Paltani, S., Courvoisier, 
        T.J.-L., et al. 1999, A\&AS, 134, 89
\bibitem[van Leer 1977]{vanl77} van Leer, B. 1977, J. Comp. Phys., 
        23, 276
\bibitem[van der Marel et al.\ 1997]{vdm97} van der Marel, R.P.,
        de Zeeuw, P.T., Rix, H.-W., \& Quinlan, G.D. 1997, Nature,
        385, 610
\bibitem[Vedder et al.\ 1988]{vtc88} Vedder, P.W., Trester, J.J., \& 
        Canizares, C.R. 1988, \apj, 332, 725
\bibitem[White \& Sarazin 1991]{ws91} White, R.E.III, \& Sarazin, C.L. 1991,
        \apj, 367, 476
\end{thebibliography}
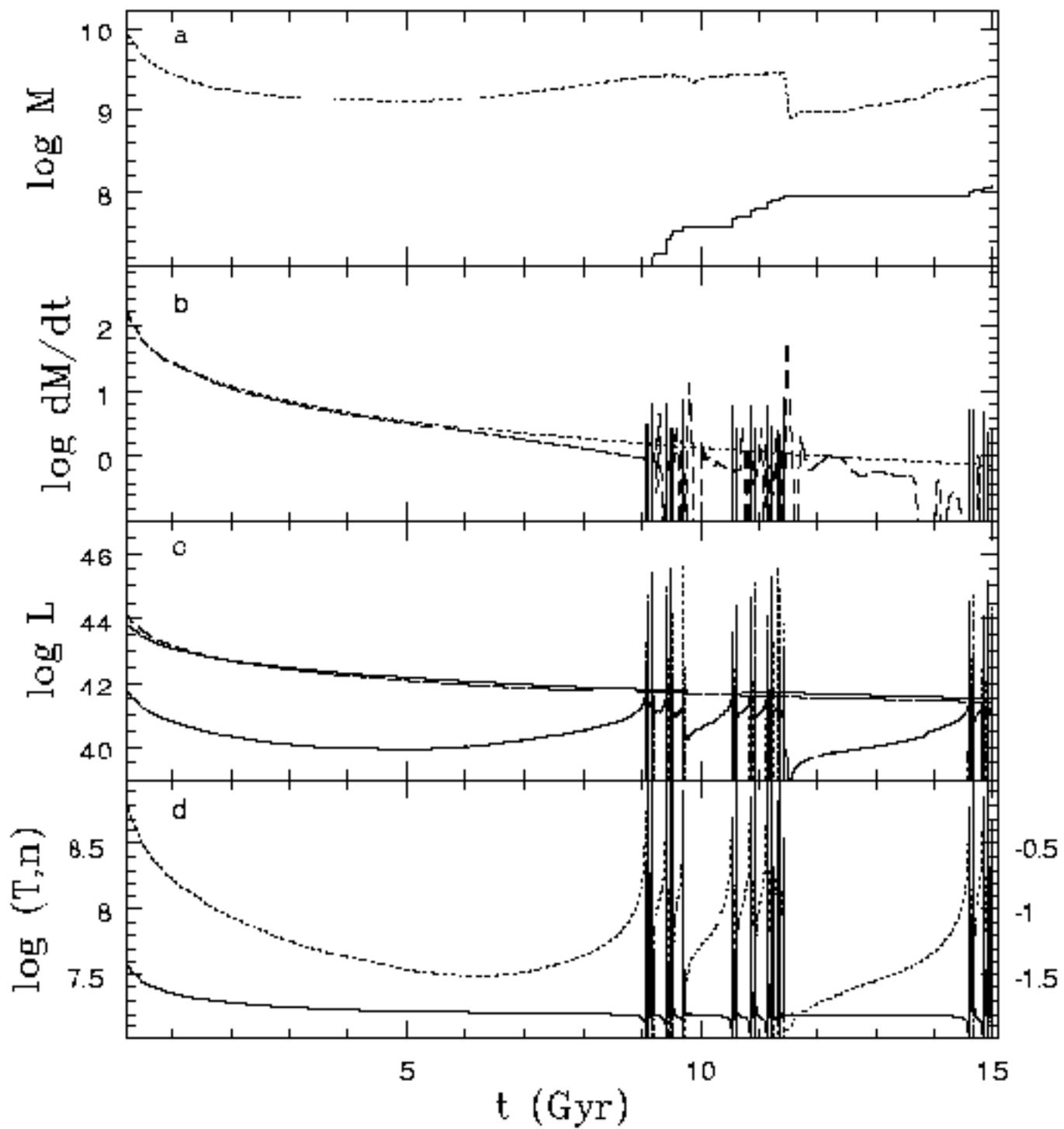
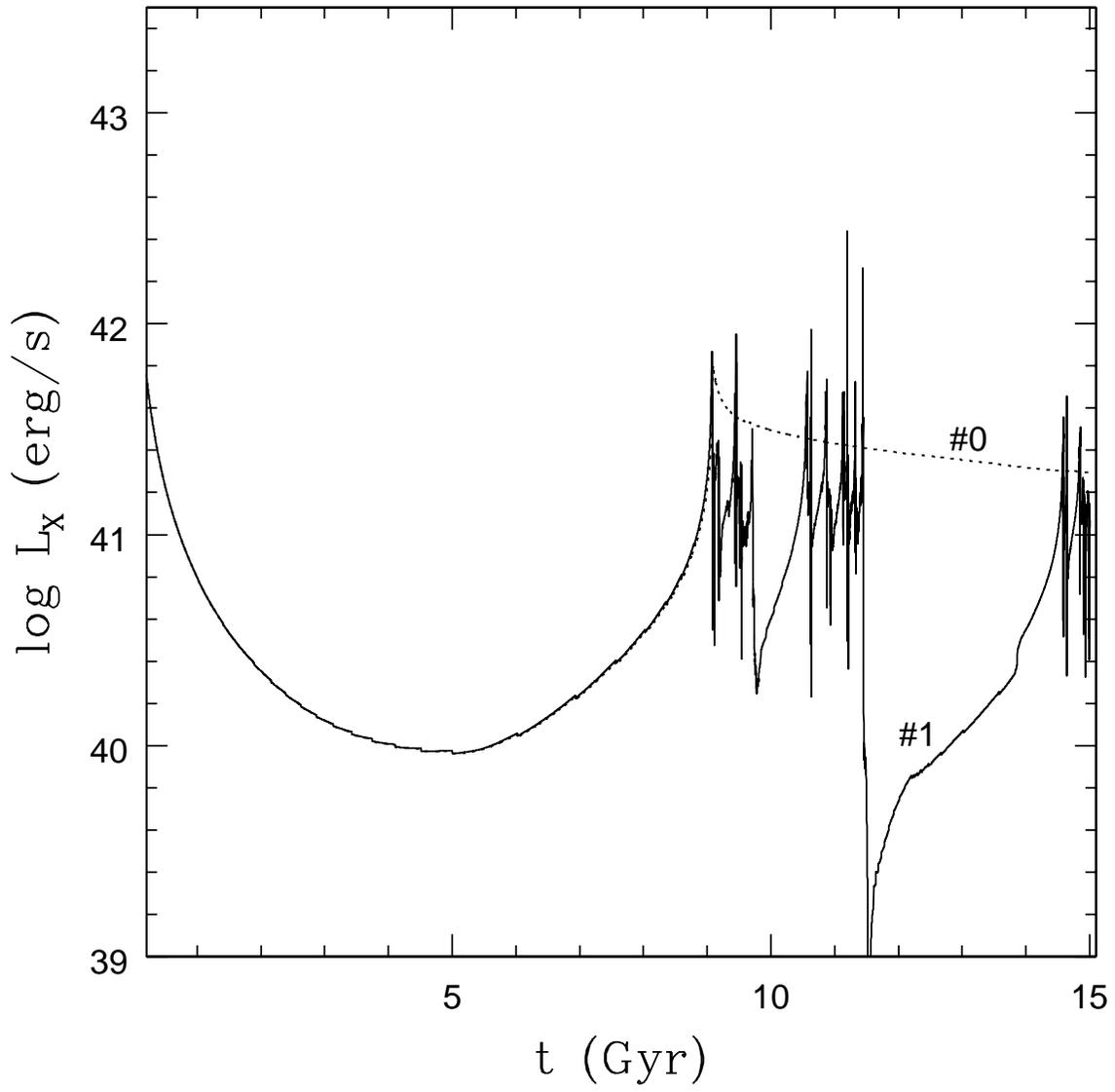
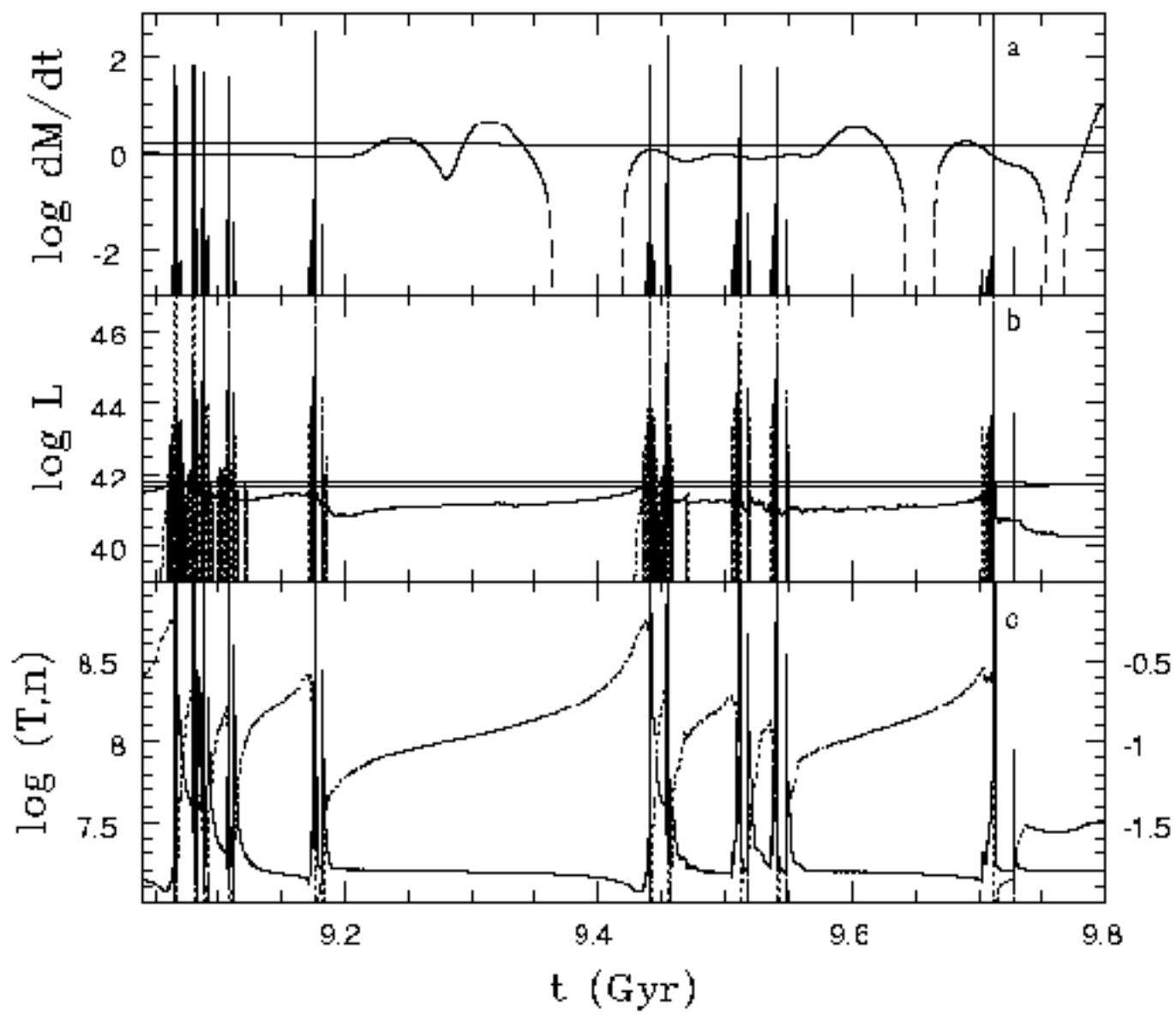
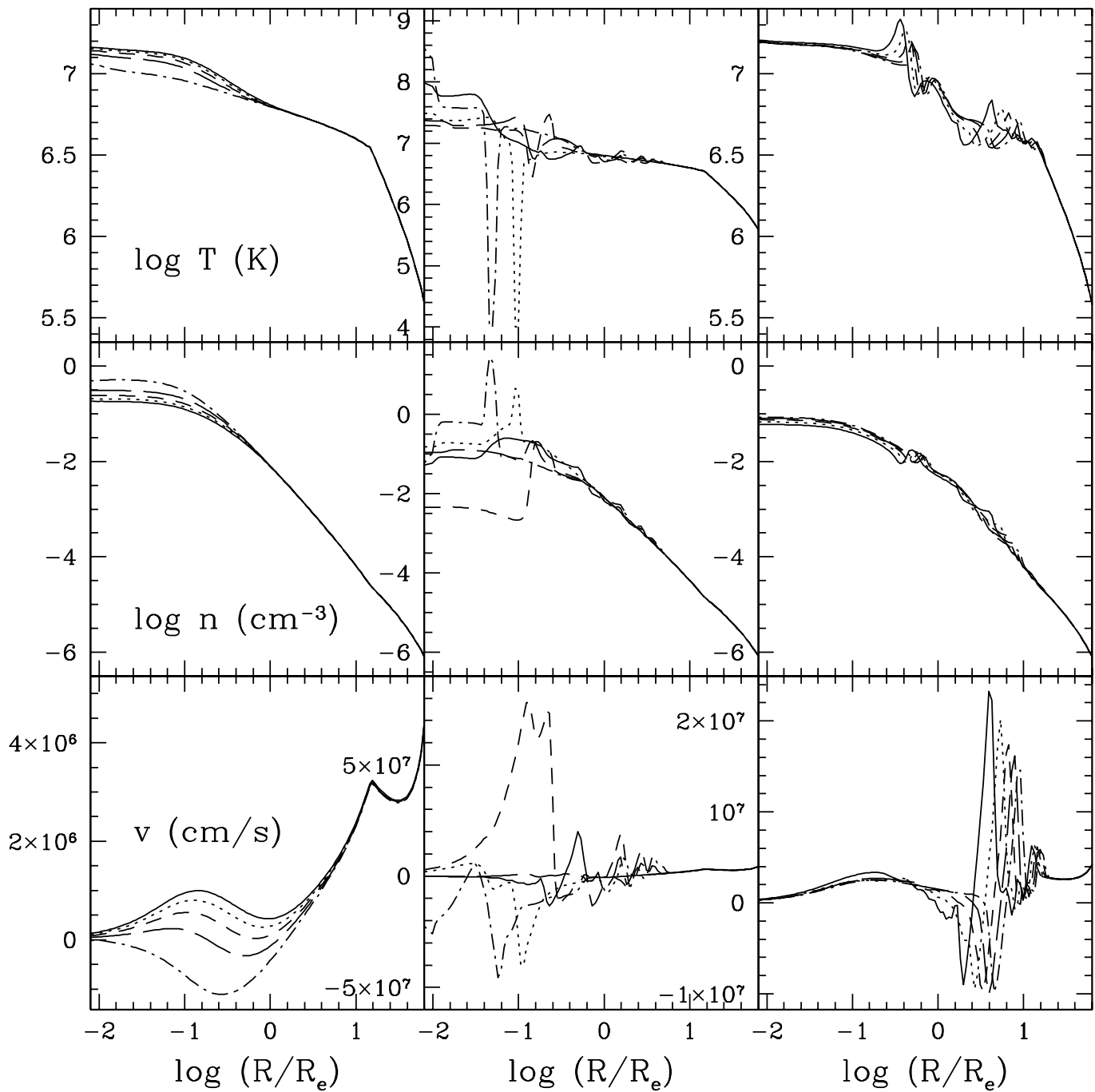
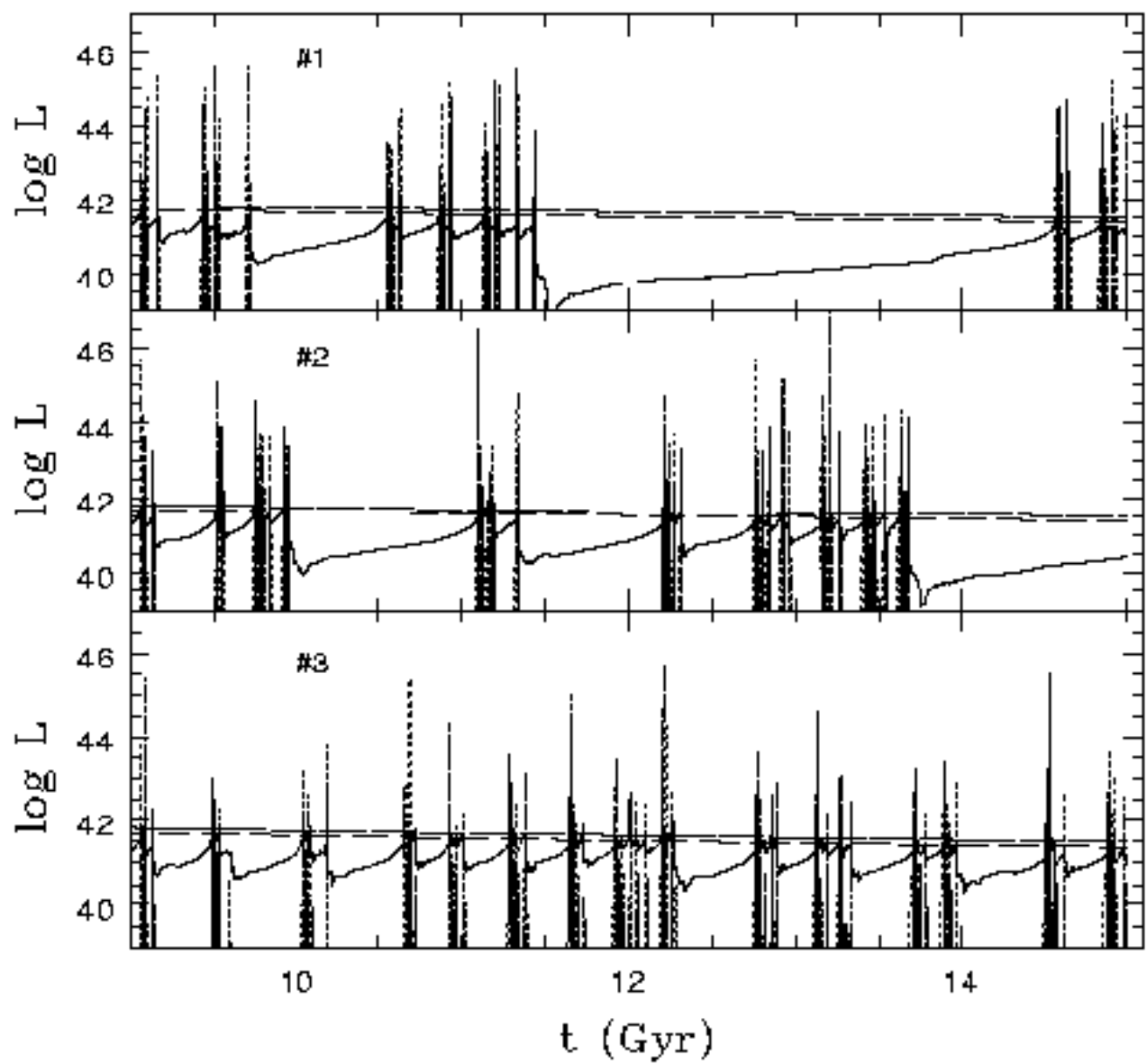
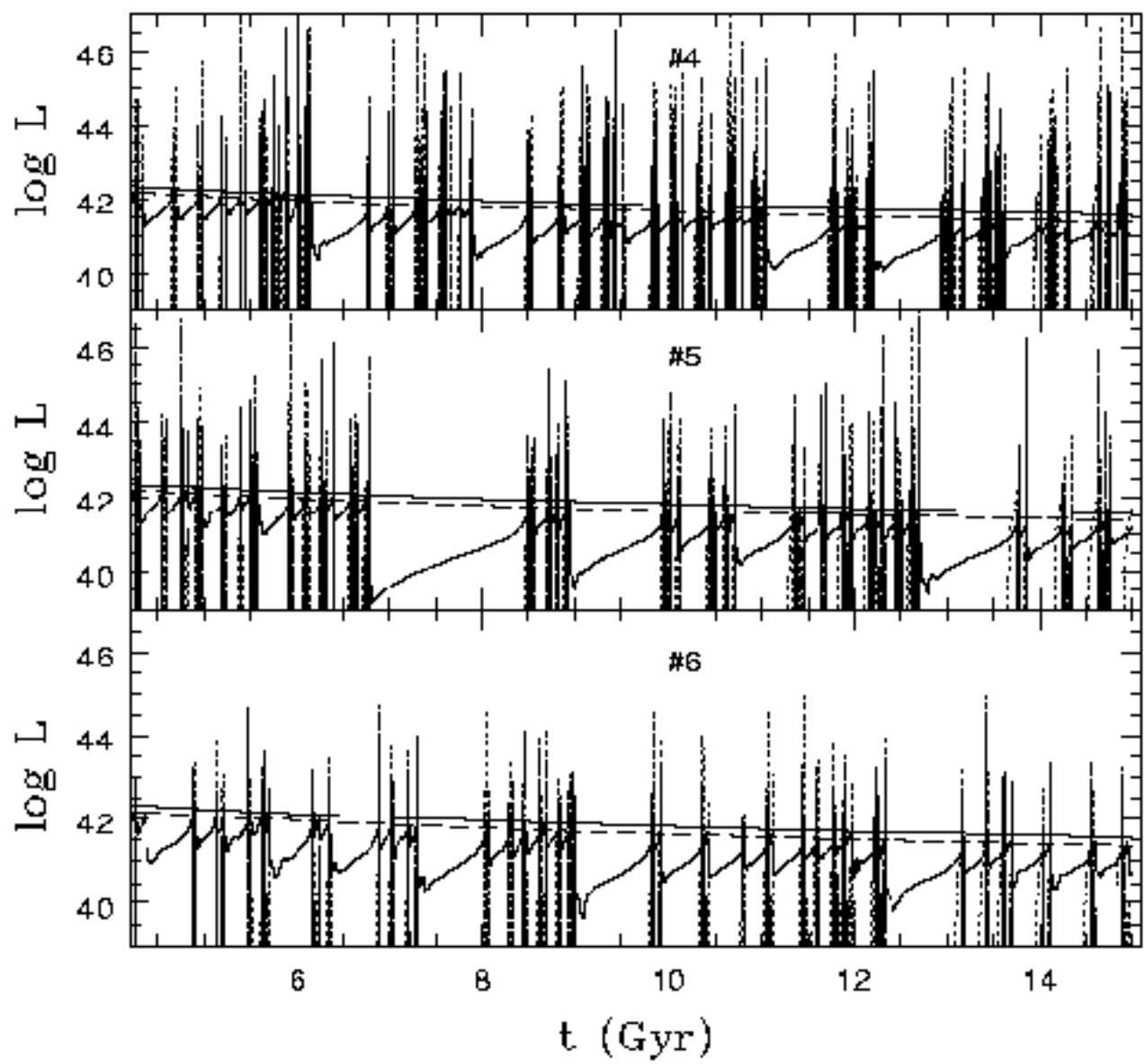
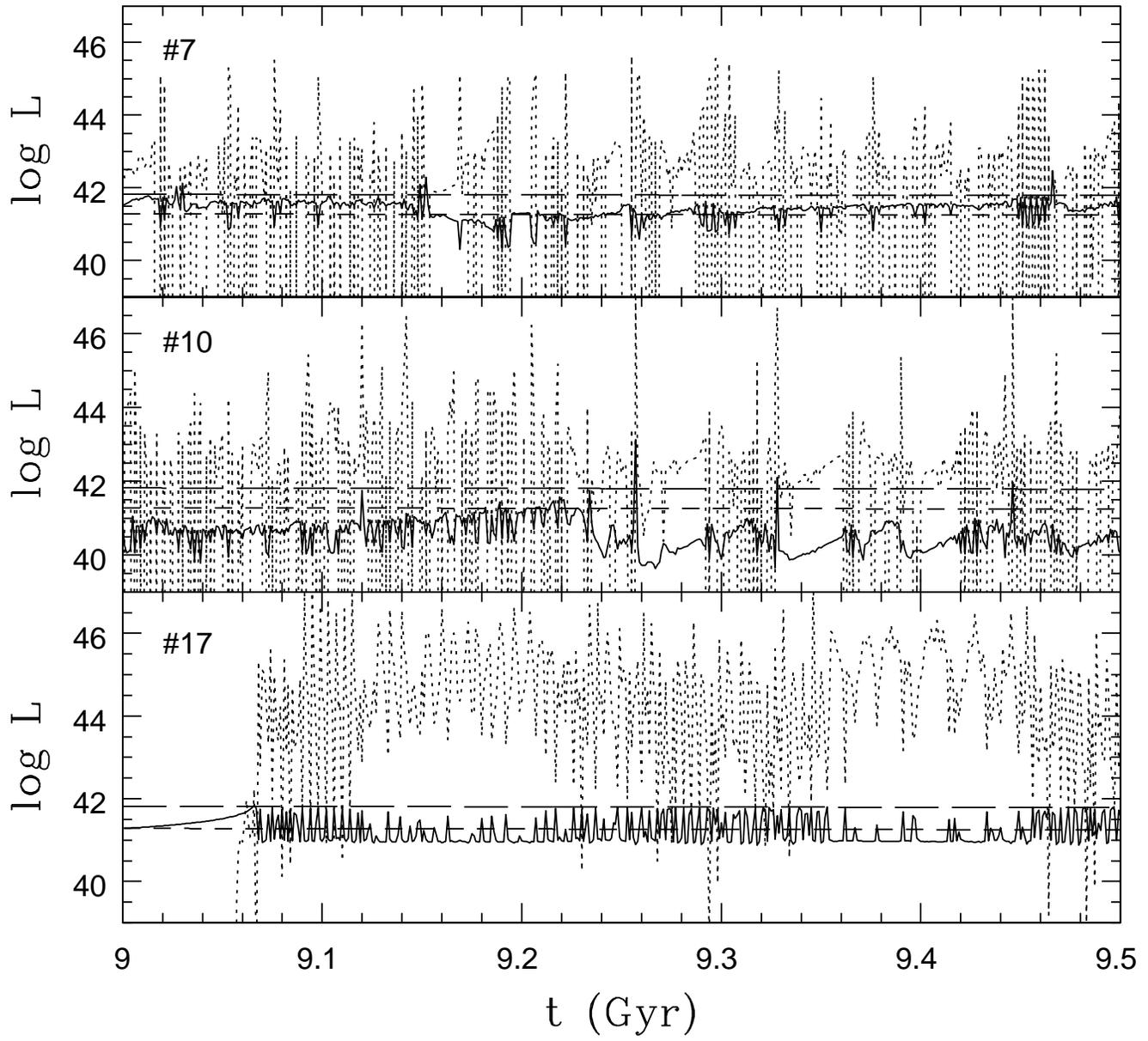
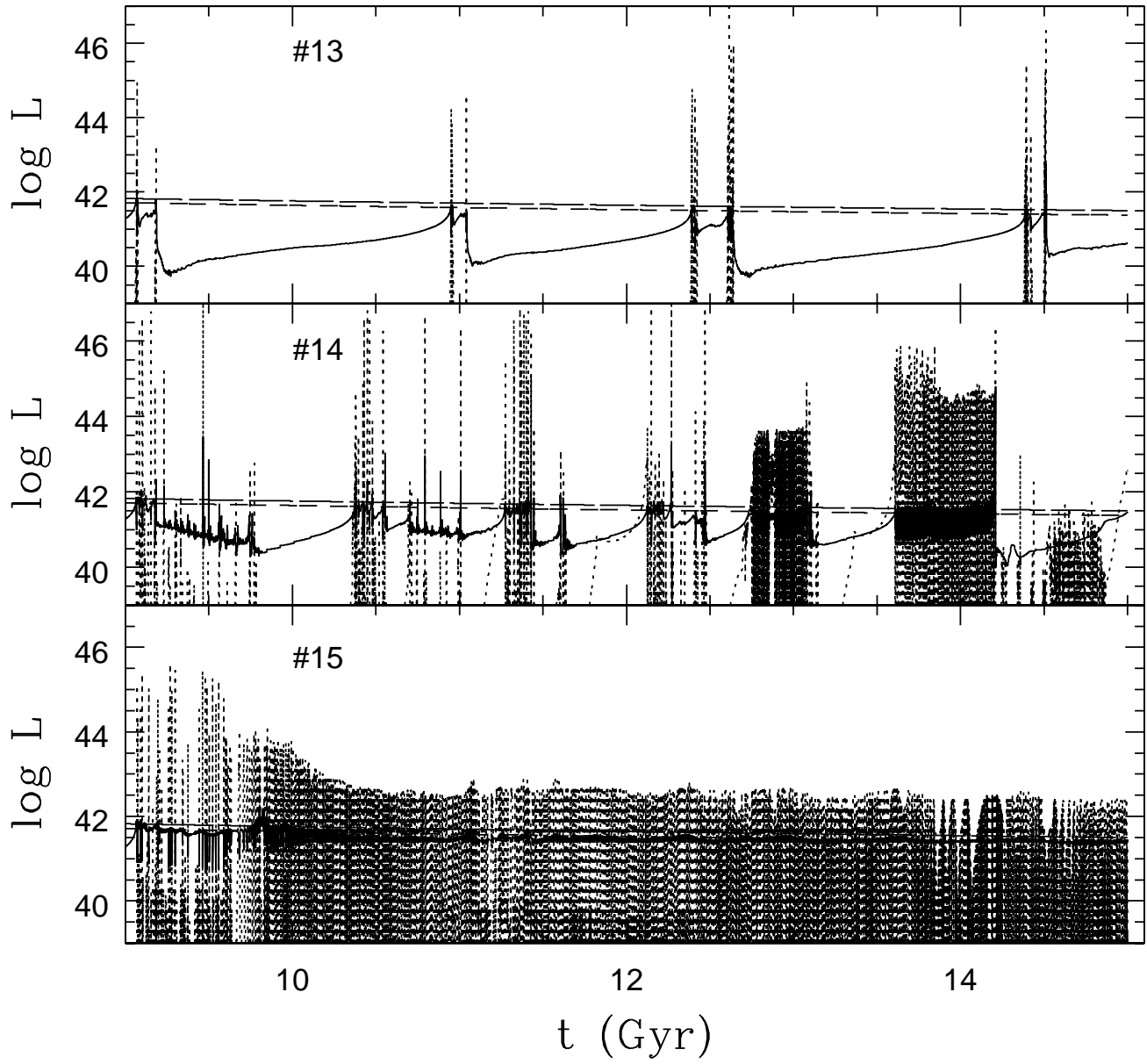
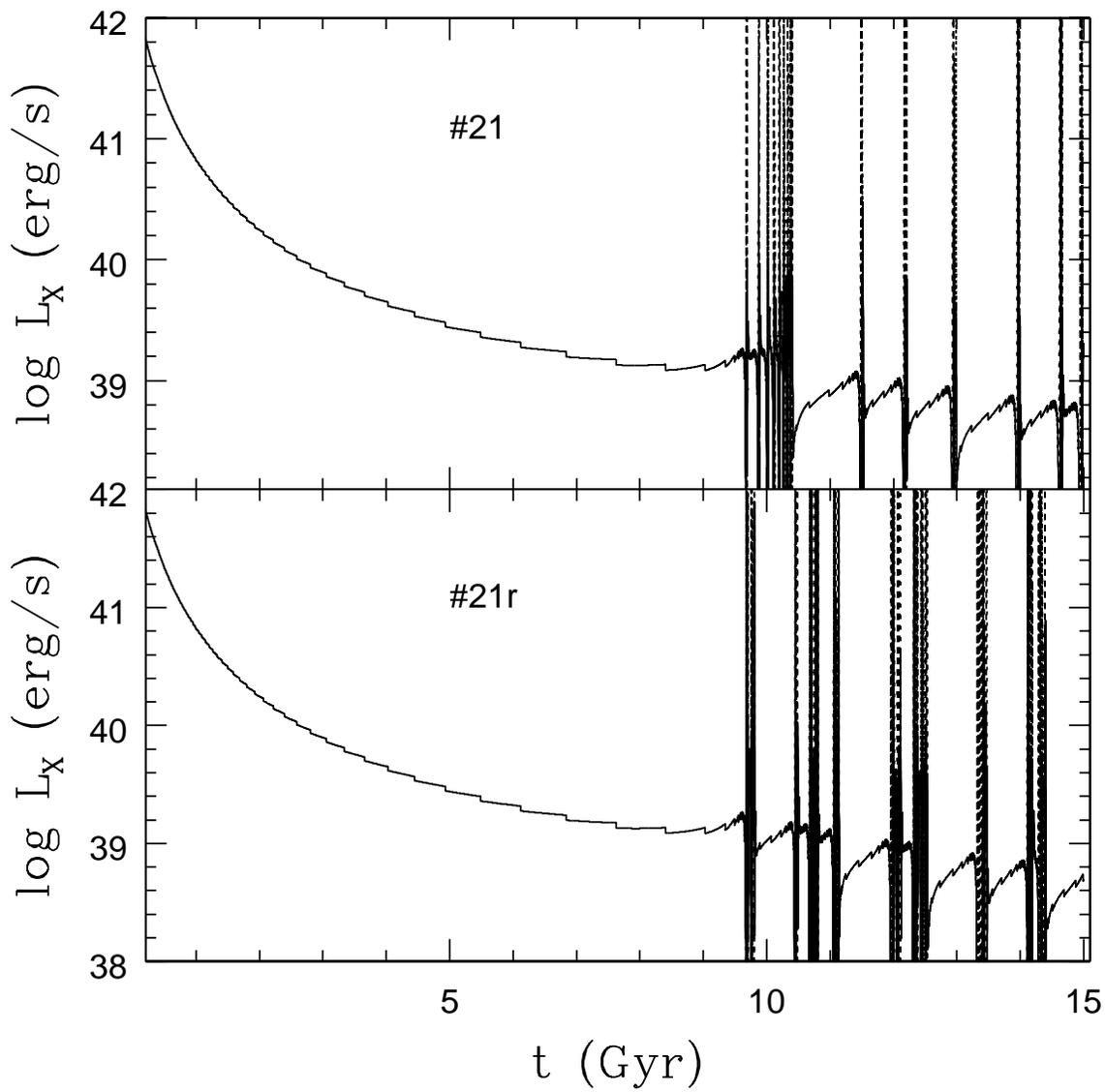
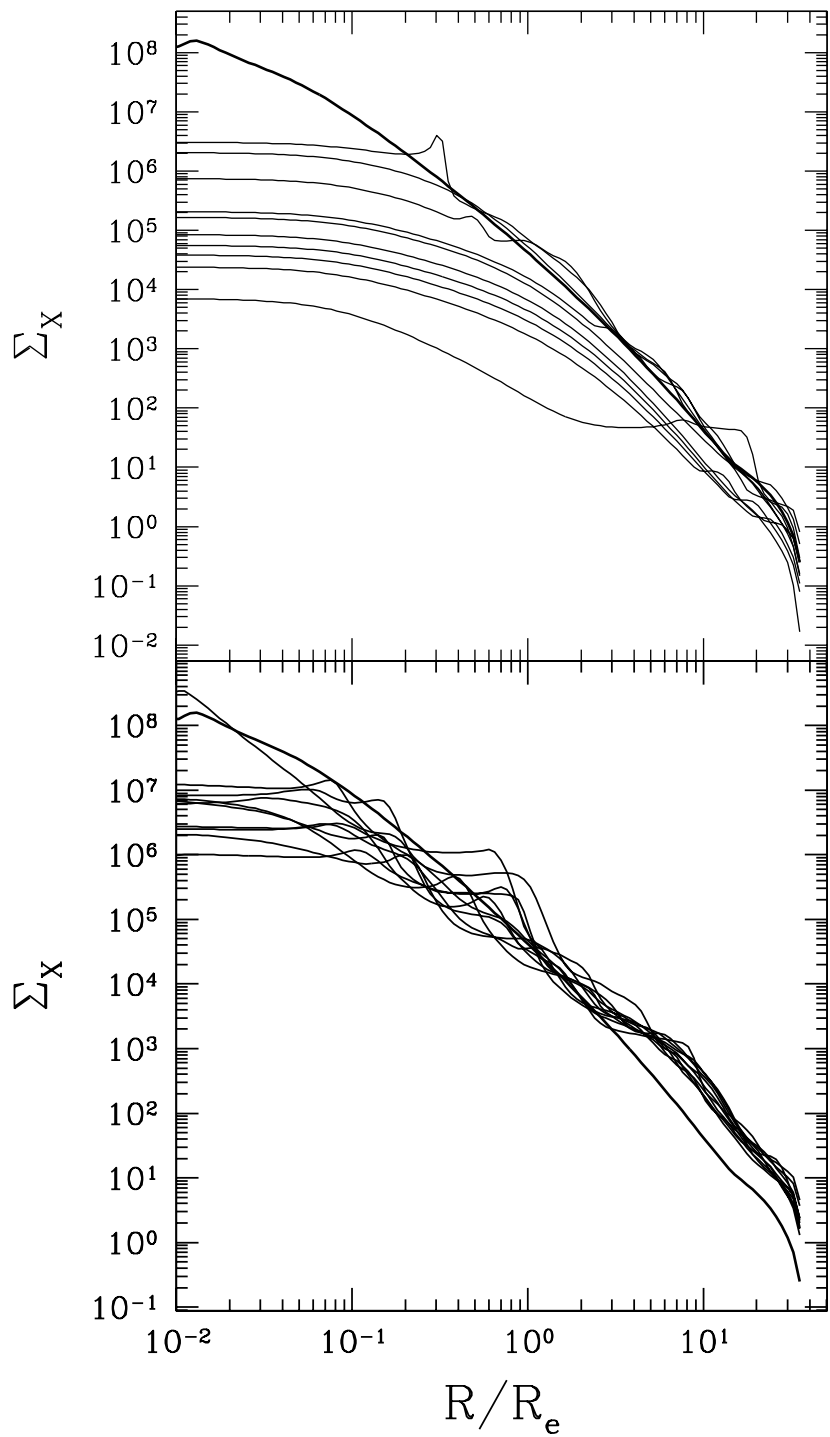
\end{document}